\documentclass{iopart}

\usepackage{graphicx}
\usepackage[colorlinks=true, citecolor=red, urlcolor=Orange, linkcolor=blue]{hyperref}
\usepackage{iopams}
\usepackage[usenames,dvipsnames]{color}
\usepackage{bm}
\usepackage[small,bf]{caption}
\usepackage{tabularx}

\begin{document}
\bibliographystyle{unsrt}

\title{Quantum Limits of Interferometer Topologies for Gravitational Radiation Detection}

\author{Haixing Miao, Huan Yang, Rana X Adhikari, and Yanbei Chen}
\address{Division of Physics, Math, and Astronomy, California Institute of Technology, Pasadena, California, 91125}
\ead{rana@caltech.edu}

\begin{abstract}
In order to expand the astrophysical reach of gravitational wave detectors, several interferometer
topologies have been proposed, in the past, to evade the thermodynamic and quantum mechanical limits
in future detectors. In this work,
we make a systematic comparison among these topologies by considering their sensitivities and complexities.
We numerically optimize their sensitivities by introducing a cost function that tries to
maximize the broadband improvement over the sensitivity of current detectors. We find
that frequency-dependent squeezed-light injection with a hundred-meter scale filter cavity
yields a good broadband sensitivity, with low complexity, and good robustness against optical loss.
This study gives us a guideline for the near-term experimental research programs in enhancing the
performance of future gravitational-wave detectors.
\end{abstract}

\pacs{04.80.Nn, 07.05.Dz, 07.10.Fq, 07.60.Ly, 95.55.Ym}

\submitto{\CQG}

\maketitle

% TODO
%
%
%
%

\newpage
\section{Introduction and Summary}
\label{s:Intro}
Over the past two decades, an international array of laser-interferometer gravitational-wave (GW)
detectors has been built and operated near their theoretical sensitivity
limits~\cite{PF:RPP2009, Hartmut:2010, Virgo:2011, TAMA:2009}.  No direct detection
of gravitational waves has yet been made and this is consistent with the low event rates predicted
by our knowledge of astrophysics~\cite{LSC:rates}.

The 2$^{\rm nd}$ generation of detectors, which are now being
assembled (Advanced LIGO\,\cite{aLIGO:web}, Advanced Virgo\,\cite{aVirgo:TDR},
GEO-HF\,\cite{Harald:2010un} and KAGRA\,\cite{Somiya:2011tb}), are expected to improve
sensitivities by a factor of $\sim$10 compared with the
first generation, and are expected to make direct detections.
In order to move from the era of detections to the era of precision GW measurement, the detector
sensitivities must be further improved\,\cite{Rana:RMP2014}.

The sensitivity of a laser interferometer gravitational-wave detector is limited by many noise
sources.  Among them, the quantum noise is due to ground-state fluctuations of the electromagnetic
field,  which beat with the laser field to produce shot noise and radiation pressure noise.
For 2$^{\rm nd}$ generation detectors, quantum noise is dominant in a large fraction of the entire
observation band.  Furthermore, the quantum noise is at a level
at which the Heisenberg Uncertainty Principle for kilogram scale masses becomes important.

Although many sources of noise can be regarded as having entered interferometer
output data through ``imperfections'' of the interferometer, quantum noise is so tightly coupled
into the gravitational wave transduction process,
that improving the quantum noise often requires changing the optical configuration of the
interferometer. In the past decades, several types of strategies for improving the optical
configuration have been proposed within the community\,\cite{Corbitt:Review2004,Yanbei:Review}:

\begin{enumerate}
\item injection of squeezed light\cite{DMCLam:2010, mcclelland2011advanced, H1:Squeezing} from the interferometer's dark port
\item inserting optical filters at the interferometer's input or output port
\item reshaping the interferometer's optical transfer function in the frequency domain
\item modifying the test masses' mechanical transfer function, e.g., by using the optical spring effect associated with detuned signal recycling
\item injecting multiple carrier fields
\end{enumerate}
These strategies are meant to be combined with each other in order to synthesize
an optimal optical configuration. In this paper, regarding (i) above,
we shall always assume that squeezed light will be injected; for (ii), depending
on whether we use input or output filters, we will consider two options:
\begin{itemize}
\item {\it frequency dependent squeeze angle}---injecting squeezed light with an optical filter
  cavity\,\cite{KLMTV2001, HaEA2003, Corbitt2004, Khalili2010, PhysRevD.88.022002};
\item {\it frequency dependent readout}---filtering the output with a cavity to measure appropriate
           signal quadratures at different frequencies\,\cite{Vyatchanin1995, KLMTV2001, Kha2007}.
 \end{itemize}
Next, for each of the two options above for modifying the input-output optics, we will consider
 one of the following four options for the interferometer itself:
\begin{itemize}
  \item keeping the signal-recycled configuration of Advanced LIGO [uses strategy (iv) above];
  \item {\it speed-meter configurations} \cite{Khalili2, Pur2002, PuCh2002, Danilishin2004, Chen2011}, i.e.,
    measuring a quantity that is proportional to the test mass speed at low frequencies, by inserting
    an additional long cavity [using strategy (iii) above];
  \item {\it long signal recycling cavity}---using a km scale signal recycling cavity to have a
    frequency-dependent response [using strategy (iii) and (iv)];
  \item {\it dual-carrier scheme} \cite{Yanbei:Local}---introducing an additional laser field to gain
    another readout channel; in particular, we shall also consider the so-called
    {\it local-readout scheme},  in which the additional field is anti-resonant in the arm cavity and
    resonant in the power-recycling cavity [using strategy (iii) and (iv)].
\end{itemize}

When trying to evaluate these configurations, we will include the effect of
realistic optical losses and quantitatively compare these configurations
against a few baseline interferometer configuration (which includes realistic
levels of non-quantum noise).
This differs from the proposed Einstein Telescope (ET)\,\cite{ET2011} which assumes significant 
reduction of the non-quantum noises at low frequencies and entirely new infrastructure with 
much longer arms, we are focusing on near-term upgrades to current detectors using technologies 
that can be deployed within the existing facilities.

We will numerically optimize the sensitivity of different configurations for this next generation
detector (which we call LIGO3), with the following cost function:
\begin{equation}
{\cal C}(\bm x)=\left\{\int_{f_{\rm min}}^{f_{\rm max}} {\rm d}(\log_{10} f) \,
                          \log_{10} \left[\frac{h_{\rm aLIGO}}{h_{\rm LIGO3}(\bm x)}\right]\right\}^{-1}.
  \label{eq:cost_function}
\end{equation}
Here $[f_{\rm min}, f_{\rm max}] = [4\,{\rm Hz}, 4040\,{\rm Hz}]$ is the frequency span for
the optimization; $\bm x$ are the set of parameters of the optical configuration that we optimize
over; $h_{\rm aLIGO}$ is the square root of the total noise power spectral density of the baseline design
of Advanced LIGO (aLIGO); and $h_{\rm LIGO3}$ is the square root of the total noise power spectral
density of interferometers with various improved optical configurations. Notice that the
integration variable is $\log_{10} f$ instead of $f$, which means that we want to maximize the
improvement over aLIGO in the log-log scale.

\begin{figure}[h]
\centering
\includegraphics[width=\columnwidth]{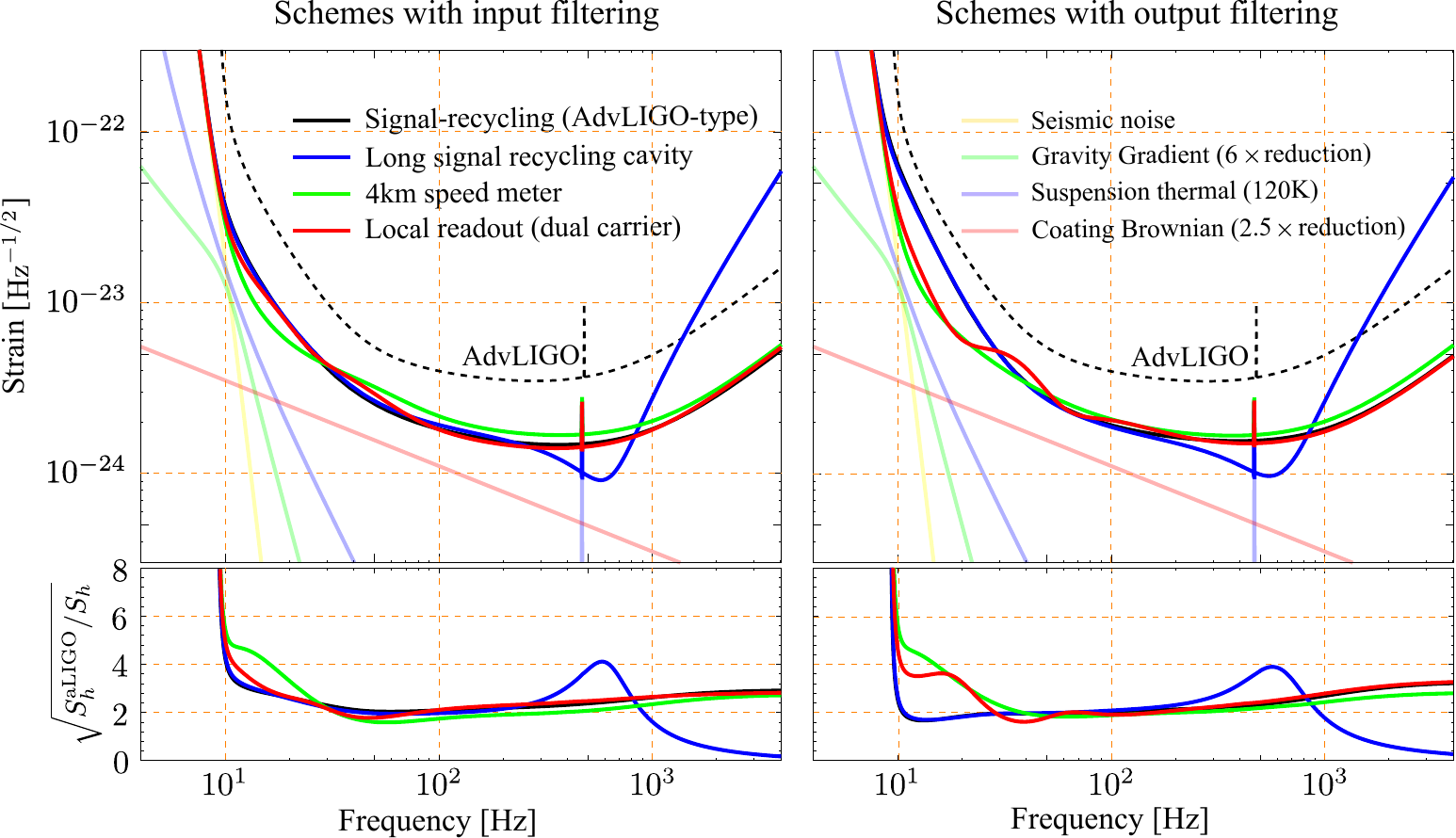}
   \caption{The optimized total noise spectrum for different schemes assuming a moderate improvement
     of the thermal noise compared with aLIGO baseline design. The left panel shows the case for schemes
     with input filtering, i.e., frequency-dependent squeezing, while the right panel shows the case for
     schemes with output filtering, i.e., frequency dependent readout, which is also called variational readout in the literature.
     The lower panels show the linear strain sensitivity improvement over Advanced LIGO.}
   \label{fig:opt_result}
\end{figure}

\begin{figure}[h]
\centering
\includegraphics[width=\columnwidth]{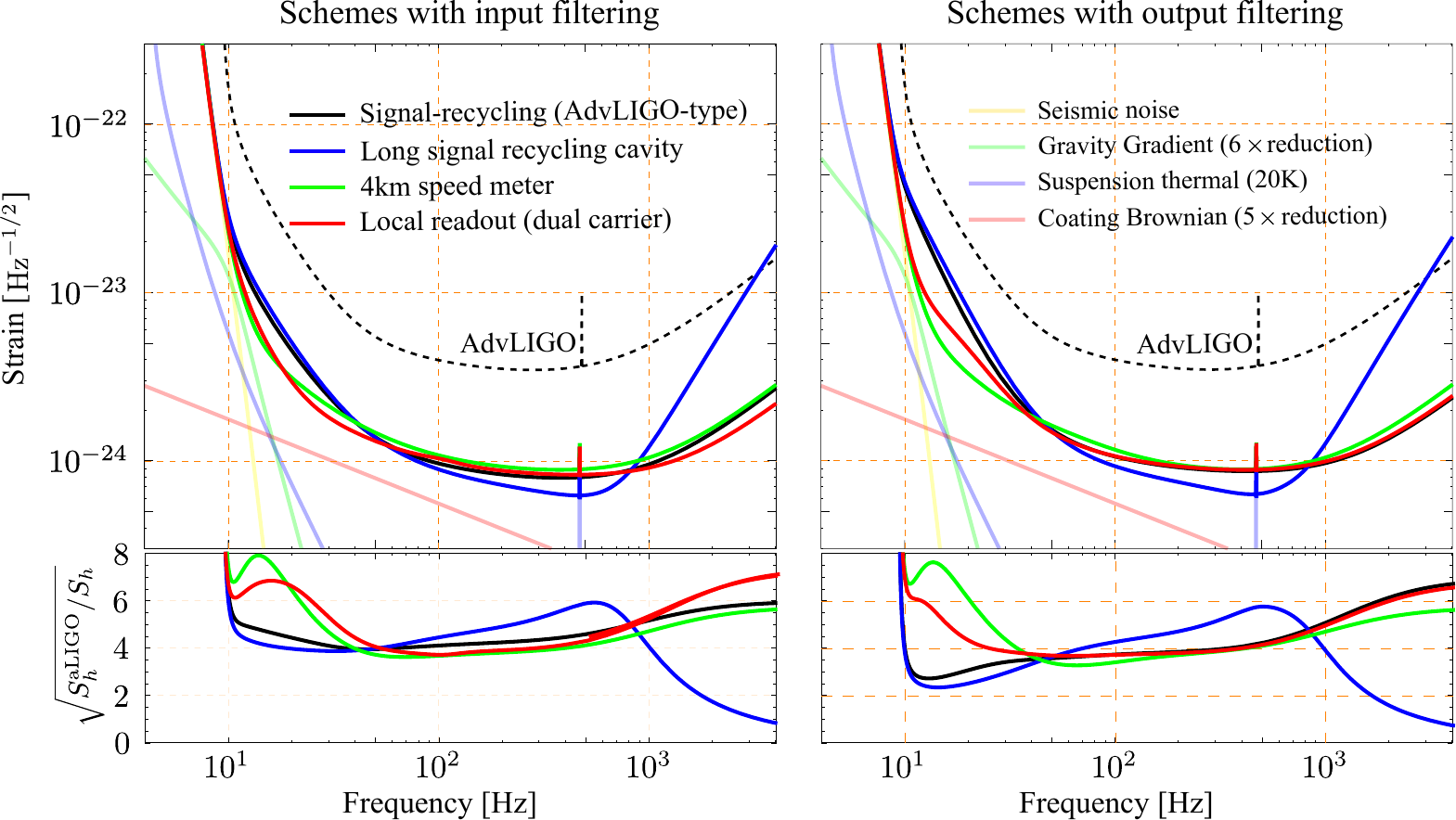}
   \caption{Optimization results for different schemes assuming more substantial thermal noise
     improvements, increasing the mirror mass from 40 -- 150\,kg, and increasing the arm cavity
     power from 800 -- 3000\,kW.}
   \label{fig:opt_result2}
\end{figure}

The results of the numerical optimization are shown in Figures\,\ref{fig:opt_result} and \,\ref{fig:opt_result2},
where we plot the total noise spectra (the quantum noise + the classical noises) for different
configurations with frequency dependent squeezing (input filtering) and frequency dependent
readout (output filtering), respectively.
In producing Figure\,\ref{fig:opt_result}, we assume a moderate
reduction in the thermal noise and the same mass and optical power as those for aLIGO.
In producing Figure\,\ref{fig:opt_result2}, we assume a more optimistic reduction in the
thermal noise, the mirror mass to be 150\,kg and the maximum arm cavity power to be 3\,MW. As we can
see, by adding just one filter cavity to the signal-recycled interferometer (the baseline aLIGO topology),
we can already obtain a substantial broadband improvement over aLIGO. Further low-frequency
enhancement can be achieved by applying either the speed meter or the local-readout
(dual-carrier Michelson) scheme.

The outline of this paper goes as follows: in Section\,\ref{s:Basics}, we summarize the basics of our quantum noise calculations;
in Section\,\ref{s:Configurations}, we introduce the interferometer topologies and the
features of the configurations that we compare;
in Section\,\ref{s:Optimize}, we introduce classical noise models for 3$^{\rm rd}$~generation
detectors and then compare different optical configurations by optimizing their parameters
under the same cost function defined in Equation\,\ref{eq:cost_function};
in Section\,\ref{s:Conclusion}, we summarize our main results.
In the appendices, we present a table of the optimized interferometer parameters, describe the non-quantum noise sources,
and also define the variables used here in comparison to the previous literature on this topic.

\newpage
\section{Basics of Quantum Noise}
\label{s:Basics}
In this section, we will briefly review the basics for evaluating quantum noise in
a laser-interferometer gravitational-wave detector by using an input-output formalism.
Additionally, we will discuss the principle behind the use of filter cavities for reducing
the quantum noise. For more detail, one can refer to a recent review
article~\cite{lrr-2012-5}.

\subsection{Input-Output Formalism}
\begin{figure}[h]
\centering
    \includegraphics[width=\columnwidth]{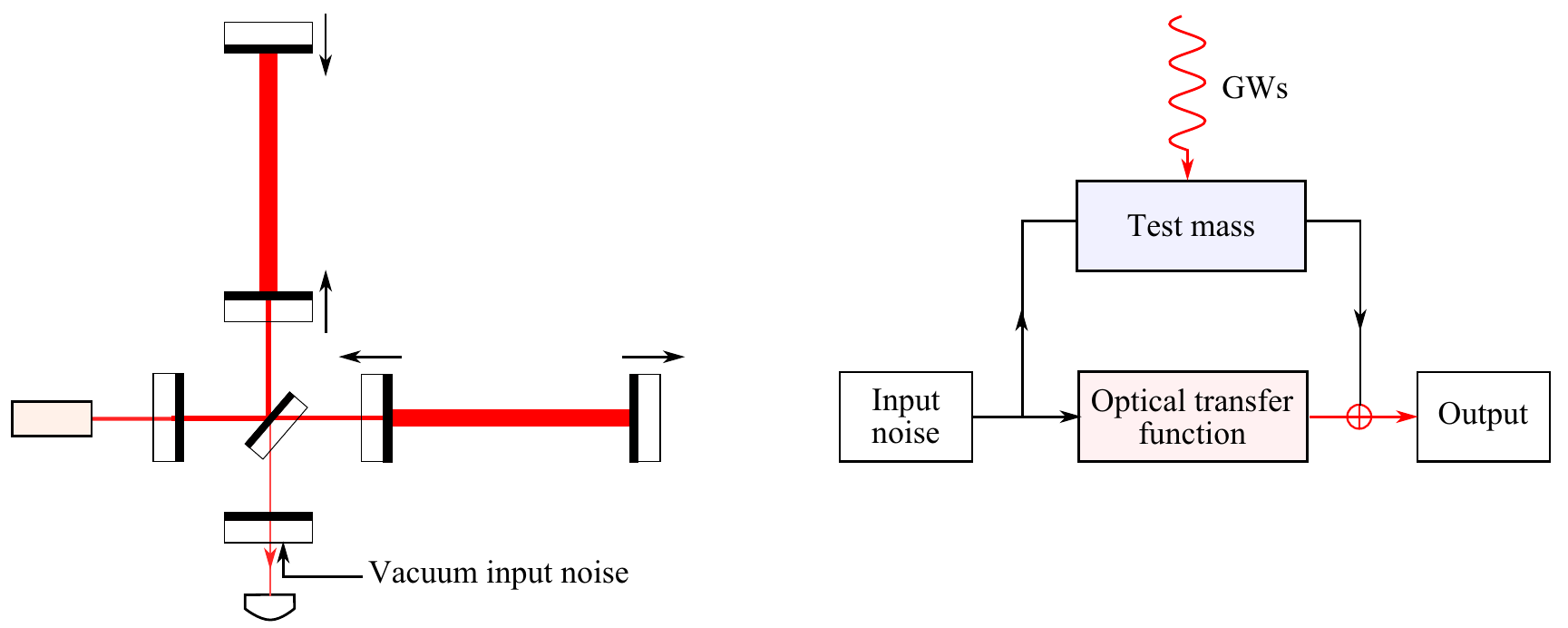}
     \caption{Schematics showing the configuration of an interferometric gravitational-wave (GW)
              detector (left) and the block diagram illustrating the two paths (upper and lower) through which the
              vacuum fluctuation propagates to the output (right).}
     \label{fig:config}
\end{figure}

When analyzing the quantum noise of a laser interferometer, as shown schematically in
Figure\,\ref{fig:config}, we assume linearity and stationarity of the system; a frequency-domain analysis
can therefore be applied with the noise and signal
propagating through the system via various linear transfer functions. There are two types of
noise: (i) the shot noise, also called the readout noise, is the one that comes from the
measurement device itself---in the context here, arising from the phase fluctuation of the light,
and it usually decreases as we increase the measurement strength (the optical power). Its propagation
is denoted by the lower path of the block diagram in Figure\,\ref{fig:config} (ii) the back-action noise,
also called the radiation-pressure noise here, is the one that disturbs the test mass due to noise
in the device, and it usually increases when the measurement strength increases. Its propagation
is shown by the upper path of the diagram. In general, these two types
of noise are mixed with each other. To evaluate detector sensitivity, the key is then to analyze
how the noise and signal propagate and to identify those transfer functions, which gives the
input-output relations.

For these interferometers, the photocurrent output $I_{\rm out}$ that we measure is linearly
proportional to a certain optical quadrature---a linear combination of the amplitude
quadrature $b_1$ and phase quadrature
$b_2$\footnote{These quadratures are related to the upper $b(\Omega)$ and lower audio sideband
$b(-\Omega)$ via $b_1=[b(\Omega)+b^{\dag}(-\Omega)]/\sqrt{2}$ and
$b_2=[b(\Omega)-b^{\dag}(-\Omega)]/(i\sqrt{2})$.}:
\begin{equation}
I_{\rm out}(\Omega)\propto b_1(\Omega)\sin\zeta +b_2(\Omega)\cos\zeta,
\end{equation}
where $\zeta$ is the readout quadrature angle and can be adjusted by the phase of the
local oscillator (the optical field that beats with the interferometer output). In terms of amplitude
and phase quadratures, the input-output relation can generally be put into the following form:
\begin{equation}
\left[\begin{array}{c}b_1(\Omega)\\b_2(\Omega)\end{array}\right]=\left[\begin{array}{cc}M_{11}
(\Omega)&M_{12}(\Omega)\\M_{21}(\Omega)&M_{22}(\Omega)\end{array}\right]
\left[\begin{array}{c}a_1(\Omega)\\a_2(\Omega)\end{array}\right]+
\left[\begin{array}{c}v_1(\Omega)\\v_2(\Omega)\end{array}\right]h(\Omega).
\end{equation}
Here $\Omega=2\pi f$ is the angular frequency; $b_1 (a_1)$ and $b_2 (a_2)$ are the output (input)
amplitude quadrature and phase quadrature, respectively; $M_{ij}$ are the elements of the transfer
matrix, which depend on the specific optical configuration; $v_i$ quantify the detector response to
the gravitational-wave strain $h$. More compactly, one can rewrite the input-output relation in a vector form: $
\bm b(\Omega) = {\bf M}(\Omega) \bm a(\Omega)+ \bm v(\Omega) h(\Omega)\,.$
Different configurations will have different transfer matrices and
response functions to the gravitational-wave signal---thus different input-output relations. In the
following sections, we will see an interesting assortment of them. 

Given the above input-output relation and
the readout angle $\zeta$ of the homodyne detection by adjusting the local oscillator phase, the output current of the photodiode is proportional to $y(\Omega)=\bm d_{\zeta}^{\rm T} \bm b(\Omega)$, namely, 
\begin{equation}\label{eq:out}
y(\Omega)= \bm d_{\zeta}^{\rm T} {\bf M}(\Omega) \bm a(\Omega) + \bm d_{\zeta}^{\rm T} \bm v(\Omega)h(\Omega)\,,
\end{equation}
where the readout vector is defined as $\bm d_{\zeta}\equiv (\sin\zeta,\,\cos\zeta)^{\rm T}$. The first term is the quantum noise, while the second term is the output response to the GW signal. The detector sensitivity is quantified by the noise power spectral density~\footnote{The single-sided power spectral density $S_A(\Omega)$ for any quantity $A$ is
defined by [cf. Equation (22) in\,\cite{KLMTV2001}]: $\langle A(\Omega) A^{\dag}(\Omega')\rangle_{\rm sym}=\frac{1}{2}\langle A(\Omega) A^{\dag}(\Omega')+ A^{\dag}(\Omega') A(\Omega)\rangle=\frac{1}{2}2\pi S_A(\Omega)\delta(\Omega-\Omega')$.}
(normalized with respect to the GW strain $h$):
\begin{equation}
S_h(\Omega)=\frac{\bm d_{\zeta}^{\rm T}{\bf M}(\Omega)\,{\bf S}(\Omega)\,{\bf M}^{\dag}(\Omega)\bm d_{\zeta}}
{|\bm d_{\zeta}^{\rm T} {\bm v}(\Omega)|^2},
\end{equation}
where $\bf S$ is the noise spectral-density matrix for the input amplitude quadrature $a_1$ and
the phase quadrature $a_2$---$\langle a_i(\Omega) a^{\dag}_j(\Omega')\rangle_{\rm sym}\equiv
\pi\,S_{ij}(\Omega)\delta(\Omega-\Omega')\; (i, j=1, 2)$,
and in particular for non-squeezed light (vacuum) input, its elements
are $S_{11}=S_{22}=1$ and $S_{12}=S_{21}=0$ (uncorrelated amplitude and phase noise).

\begin{figure}[h]
\centering
   \includegraphics[width=0.5\columnwidth]{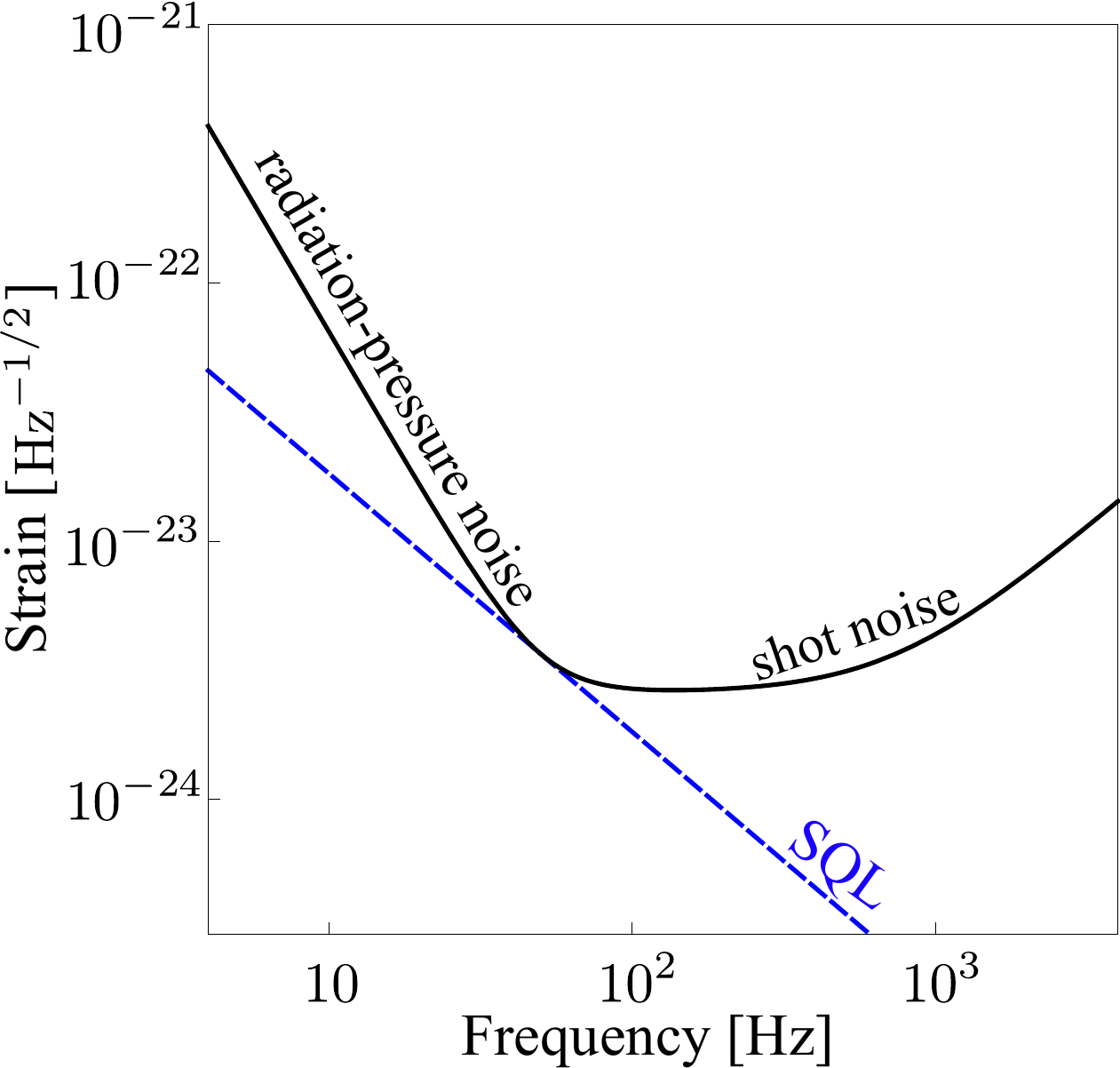}
       \caption{The quantum-noise spectral density $S_h^{1/2}$ for a broadband RSE
                interferometer given the same specification of aLIGO---$m=40$\,kg and $P_c=800$\,kW.}
   \label{fig:Sh_tuned}
\end{figure}

Taking a broadband, resonant sideband extraction (RSE), tuned dual-recycled interferometer
(the baseline configuration of aLIGO) for example, the input-output relation is
given by~\cite{KLMTV2001}:
\begin{eqnarray}
\left[\begin{array}{c}b_1\\
b_2\end{array}\right]=e^{2i\phi}\left[\begin{array}{cc}1&0\\-{\cal K}(\Omega)
&1\end{array}\right]\left[\begin{array}{c}a_1\\a_2\end{array}\right]
+e^{i\phi}\left[\begin{array}{c}0\\\sqrt{2{\cal K}(\Omega)}\end{array}\right]\frac{h(\Omega)}{h_{\rm SQL}}.
    \label{eq:io_tuned}
\end{eqnarray}
Here we have introduced:
\begin{equation}
\phi \equiv \arctan (\Omega/\gamma), \quad \mathcal{K}(\Omega) \equiv \frac{2\gamma\,\iota_c}
{\Omega^2(\Omega^2+\gamma^2)}, \quad h_{\rm SQL} \equiv \sqrt{\frac{8\hbar}{m \Omega^2 L^2}}
\end{equation}
with $\gamma$ the arm cavity bandwidth, $L$ the arm cavity length,
$\iota_c\equiv {8\omega_0 P_c}/({m L c})$ for quantifying the measurement strength,
$\omega_0$ the laser angular frequency, and $P_c$ the arm cavity power.
If we measure the phase quadrature by choosing the readout phase to be
$\zeta=0$, the corresponding noise power spectral density will be:

\begin{equation}
S_h(\Omega) = \left[{\cal K}(\Omega)+\frac{1}{{\cal K}(\Omega)} \right]\frac{h_{\rm SQL}^2}{2} \ge h_{\rm SQL}^2=\frac{8\hbar}{m \Omega^2 L^2}.
\end{equation}

The first term, proportional to the optical power (${\cal K}\propto P_c$), is the radiation-pressure noise
and comes from the fluctuation of the input amplitude quadrature $a_1$; the second term, inversely proportional to
the optical power, is the shot noise and comes from the fluctuation of the input phase
quadrature $a_2$. In this simple scenario, the sensitivity is limited by the standard quantum limit
(SQL)---the benchmark for the strength of quantum noise\,\cite{BrKh1999a}. In Figure\,\ref{fig:Sh_tuned},
we plot $S_h^{1/2}(\Omega)$---the radiation-pressure noise dominates at low frequencies and the shot
noise dominates at high frequencies.

\newpage
\section{Optical Topologies}
\label{s:Configurations}
In this section, we briefly describe the strategies for configuration improvements that will be
used in Section\,\ref{s:Optimize}. We will compute the corresponding quantum noise spectrum
using the input-output formalism introduced in Reference\,\ref{s:Basics}.

\subsection{Frequency-dependent squeezing angle (input filtering)}
\label{s:input}

\begin{figure}[ht]
\centering
    \includegraphics[width=\columnwidth]{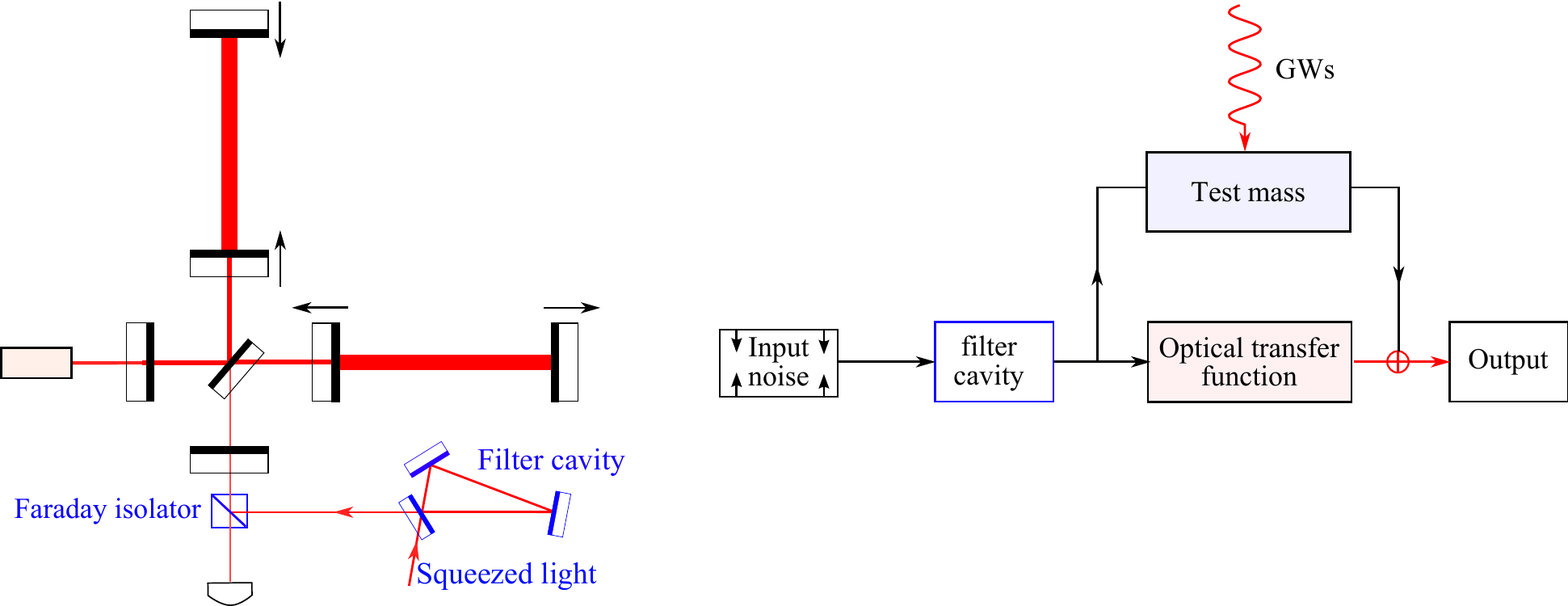}
     \caption{Schematic showing the frequency dependent squeezed light
              phase angle rotation scheme (left) and its associated block diagram (right).}
    \label{fig:input_filtering}
\end{figure}

Unlike the vacuum state, for which the cross spectral density matrix for any two
orthogonal field quadratures is the identity matrix, squeezed light has the following
cross spectral density matrix for the quadratures $b_1$ and $b_2$:
\begin{eqnarray}
\label{eq:squeezematrix}
&&\left[\begin{array}{cc} S_{11}& S_{12} \\ S_{21} & S_{22}\end{array}\right]  \nonumber\\
&=&
\left[\begin{array}{cc}\cosh 2r-\sinh 2r \cos2\varphi&  -\sinh 2r \sin2\varphi\\
-\sinh2r \sin2\varphi & \cosh2r + \sinh2r \cos2\varphi \end{array}\right]
\end{eqnarray}
where $r$ is called the squeezing factor (10~dB squeezing means that $e^{2r}=10$)
and $\varphi$ is the squeezing angle.  If we define the $\zeta$ quadrature
[see Section~\ref{s:Basics} for details] as
\begin{equation}
b_\zeta = b_1 \cos\zeta + b_2\sin\zeta\,,
\end{equation}
and if we denote by $S_{\zeta}$ the spectral density of $b_\zeta$, then
the matrix becomes
\begin{equation}
S_{\zeta} =  S_{11} \cos^2 \zeta + 2 S_{12}\cos\zeta\sin\zeta +  S_{22}\sin^2\zeta\,.
\end{equation}
In particular, $S_{\varphi} = e^{-2r}$, which means the $\varphi$-quadrature (often
referred to as the squeezed quadrature) has a fluctuation that is $e^{-r}$ the level of
vacuum (in amplitude), while the $\pi/2+\varphi$ quadrature orthogonal to $\varphi$-quadrature (often
referred to as the anti-squeezed quadrature) has $S_{\varphi} =e^{+2r}$, which
is $e^{+r}$ the level of vacuum fluctuation.

The above picture exists for each audio sideband frequency $\Omega$.  Frequency-dependent
squeezed light describes a state that has different squeeze factors and/or angles for each frequency.
As schematically shown in Figure\,\ref{fig:input_filtering}, such a frequency dependence can be
realized, for example, by injecting frequency-independent squeezed light into a Fabry-Perot cavity
with a linewidth and detuning frequency comparable to the audio frequencies of interest.
If we define $b_{1,2}$ as the quadrature fields of the output, and $a_{1,2}$
the quadratures of the input field, the cavity has an input-output relation of
\begin{equation}
\label{eq:filterrotation}
\left[\begin{array}{c} b_1\\ b_2\end{array}\right] =
e^{\frac{i(\alpha_+-\alpha_-)}{2}} \left[\begin{array}{cc} \cos \frac{\alpha_++\alpha_-}{2} & - \sin \frac{\alpha_++\alpha_-}{2} \\
 \sin  \frac{\alpha_++\alpha_-}{2} &  \cos  \frac{\alpha_++\alpha_-}{2} \end{array}\right]\left[\begin{array}{c} a_1
 \\ a_2\end{array}\right]\,,
\end{equation}
with $\alpha_{\pm}$ is defined as
\begin{equation}
e^{i\,\alpha_{\pm}}\equiv \frac{i\gamma\mp\Omega-\Delta}{i\gamma\pm\Omega-\Delta}\,,
\end{equation}
where $\Delta$ and $\gamma$ are the detuning frequency and bandwidth of the filter cavity,
respectively. As indicated by Equation\,\ref{eq:filterrotation}, the quadratures undergo
a frequency dependent rotation of $(\alpha_++\alpha_-)/2$.  This converts a frequency
independent squeezed vacuum into one with a frequency dependent squeezing angle.

With the correct frequency dependence, one can rotate the squeezing angle such that
the quantum noise is reduced by a factor $e^{2r}$ over the entire frequency band, namely
(in the case of the broadband RSE interferometer)
\begin{equation}
S^{\rm opt}_h(\Omega)=e^{-2r} \left[{\cal K}(\Omega)+\frac{1}{{\cal K}(\Omega)} \right]\frac{h_{\rm SQL}^2}{2}\,.
\label{eq:sh_sqz_in}
\end{equation}
This is the optimum performance that can be realized with frequency dependent squeezed
light injection.

\begin{figure}[h]
  \centering
     \includegraphics[width=\columnwidth]{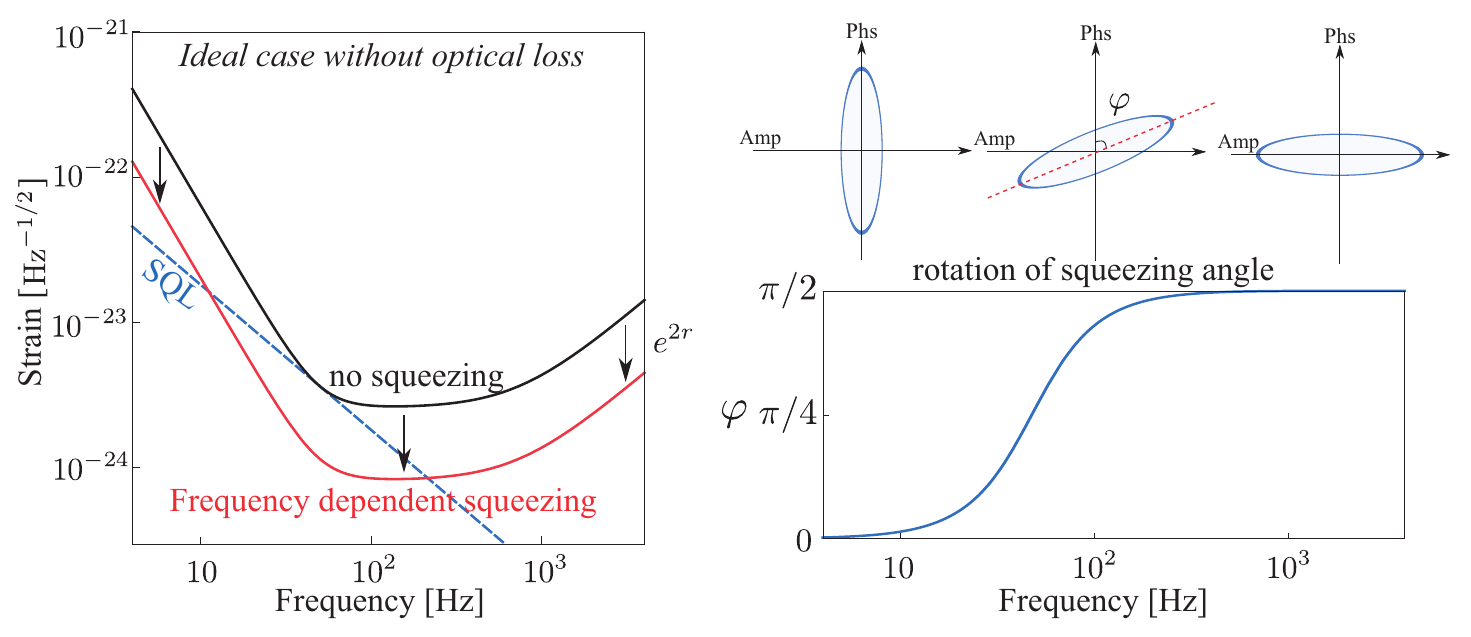}
     \caption{Noise spectrum for frequency dependent squeezing (left)
                     and illustration of the rotation of the squeezing angle (right).}
  \label{fig:Sh_input_filtering}
\end{figure}

Figure\,\ref{fig:Sh_input_filtering} shows the resulting noise spectrum in the lossless
case.  As we can see, the squeezing angle rotates in such a way that at low frequencies
the fluctuation in the amplitude quadrature is squeezed---thus reducing the
radiation-pressure noise, while at high frequencies the phase quadrature is
squeezed---thus reducing the shot noise. In order to achieve the desired rotation of
squeezing angle, the filter cavity needs to have a frequency bandwidth that is near
the frequency where the radiation pressure noise is comparable to the shot noise.

The frequency dependence of a series of such filter cavities as well as the
concomitant parameters required for realizing this frequency dependence
has been derived in\,\cite{PuCh2002}. In practice, however, the complexity of
using the ``optimal'' number of cavities and the performance degradation which
comes from optical losses, leads one to use a sub-optimal number of cavities; the
resulting degradation of the astrophysical sensitivity is negligible.

\subsubsection{The Impact of Optical Scatter Loss}
So far, we have been considering the ideal case without optical loss.
Here we provide a qualitative understanding of how loss in the filter cavity affects the sensitivity of
input filtering. Basically, the optical loss introduces additional (vacuum)
noise that is uncorrelated with the input squeezed light:
\begin{eqnarray}\label{eq:sqz_loss1}
a_1'&=&\sqrt{{\cal E}}\,n_1+\sqrt{1-{\cal E}} \, a_1,\\
a_2'&=&\sqrt{{\cal E}}\,n_2+\sqrt{1-{\cal E}} \, a_2,
\label{eq:sqz_loss2}
\end{eqnarray}
where ${\cal E}$ quantifies the total optical loss of the filter cavity and
$n_{1,2}$ are the associated noise terms in the amplitude and phase quadratures. These noise sources will
degrade the squeezing.  For example, the amplitude squeezed light originally has $S_{11}=e^{-2r}$
with $r>0$. Introducing optical loss according to Equation\,\ref{eq:sqz_loss1}, it becomes:
\begin{equation}
S_{11}'=(1-{\cal E})e^{-2r}+{\cal E}.
\end{equation}
For a completely lossy case with ${\cal E}=1$, we have $S_{11}=1$ and the squeezing simply vanishes.

\begin{figure}[h]
  \centering
  \includegraphics[width=\columnwidth]{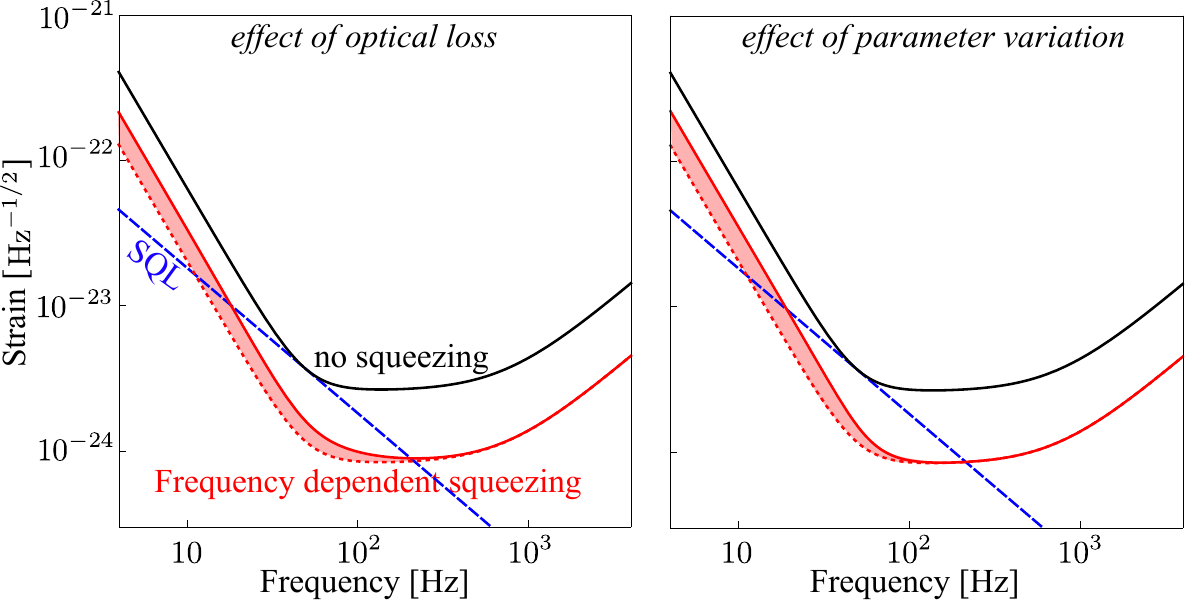}
  \caption{The effect of optical loss in the filter cavity (left) and the effect of
    parameter variation of the filter cavity (right) in the case of input filtering.
    The shaded regions illustrate the degradation of sensitivity as losses are added.
    Here we have assumed the total optical loss ${\cal E} = 20\%$ and parameter
    variations of  ${\delta T_f/T_f^{\rm opt}}={\delta \Delta_f/\Delta_f^{\rm opt}}=10\%$. }
   \label{fig:input_filtering_imperfection}
\end{figure}

The squeezed light at different audio frequencies experiences different levels of optical loss from
the filter cavity.
The low frequency part enters the cavity and circulates multiple times, while the high frequency
part barely enters the cavity. Therefore, the optical loss affects the low frequency part most
significantly (refer to \ref{s:loss} for a detailed discussion). In terms of the noise power spectrum, we
approximately have:
\begin{equation}\label{eq:loss_sqz_in}
S_h(\Omega)= \left\{[(1-{\cal E})e^{-2r}+{\cal E}]{\cal K}(\Omega)+\frac{e^{-2r}}{{\cal K}(\Omega)}\right\}
\frac{h_{\rm SQL}^2}{2},
\end{equation}
in contrast to Equation\,\ref{eq:sh_sqz_in}. Compared with the ideal frequency
dependent squeezing case, the low frequency radiation pressure noise increases due to
the optical loss and the high frequency shot noise remains almost the same.
In Figure\,\ref{fig:input_filtering_imperfection}, we show the effect of
optical loss. In producing the figure, we have assumed a total optical loss of
$20\%$, which is equivalent to a round trip loss of $40$\,ppm given a filter cavity
input mirror transmittance of $T_f=200$\,ppm.

\subsubsection{The effect of parameter variations of the filter cavity}
Apart from the optical loss, there are other imperfections of the filter cavity that
will degrade the sensitivity. In particular, here we consider the effect of variations
in the parameters of the filter cavity, which make the bandwidth $\gamma_f$, or
equivalently the input mirror transmittance $T_f$, and the detuning $\Delta_f$
deviate from their optimal value, i.e., $T_f=T_{f}^{\rm opt}+\delta T_f$ and
$\Delta_f=\Delta_f^{\rm opt}+\delta \Delta_f$.

As shown in \ref{s:tolerance}, such a parameter variation will mainly decrease
the low-frequency sensitivity
(for a reason similar to the effect of loss), and we approximately have:
\begin{equation}\label{eq:vsh_in}
S_h(\Omega)\approx S_h^{\rm opt}(\Omega)+\sinh 2r\left[\left(\frac{\delta T_f}{T_f^{\rm opt}}\right)^2+\left(\frac{\delta \Delta_f}{\Delta_f^{\rm opt}}\right)^2\right]{\cal K}(\Omega)
{h_{\rm SQL}^2}\,.
\end{equation}
If the relative error in the transmittance and the detuning can be as low as
100\,ppm, namely
$\delta T_f/T_f^{\rm opt}\sim 10^{-4}, \delta \Delta_f/\Delta_f^{\rm opt}\sim 10^{-4}$,
we have $S_h\approx S_h^{\rm opt}+ 10^{-7}{\cal K}\, h_{\rm SQL}^2$ for 10\,dB squeezing,
which is a negligible deviation from the optimal one. In
Figure\,\ref{fig:input_filtering_imperfection}, we illustrate this effect with an
exaggerated variation of $10\%$.

\subsection{Frequency dependent readout phase (output filtering)}
\label{s:output}

\begin{figure}[h]
  \centering
  \includegraphics[width=\columnwidth]{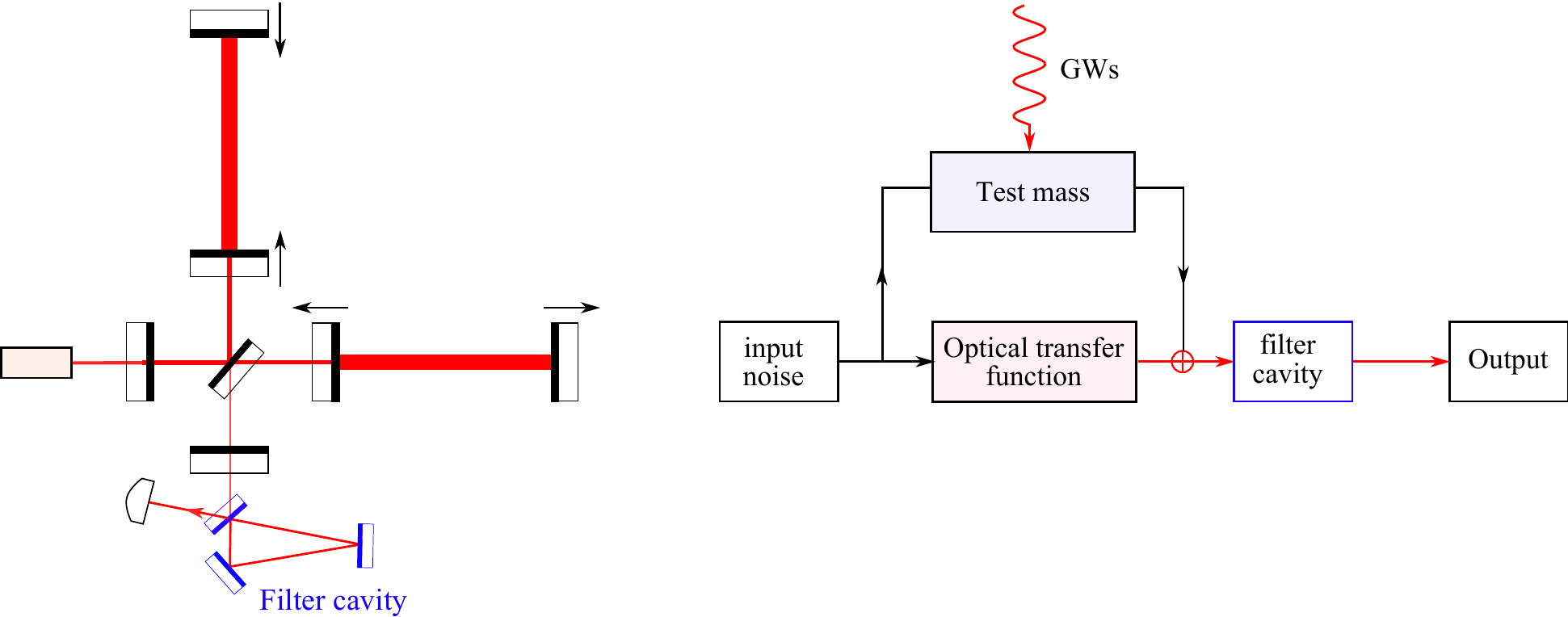}
  \caption{Schematic optical layout of the frequency dependent (or variational)
           readout scheme (left)
           and its associated block diagram (right).}
    \label{fig:output_filtering}
\end{figure}

A closely related counterpart to the frequency dependent squeezed light injection is the
frequency dependent readout angle and, as shown schematically
in Figure\,\ref{fig:output_filtering}, it uses an optical cavity to filter the detector
output allowing one to measure different optical quadratures at different frequencies.
The filter cavity has the same functionality as in the case of the frequency dependent
squeezing with the difference that it filters the outgoing fields
at the interferometer output instead of the noise fields (or squeezed light) entering from the dark port.
This scheme can be used to coherently cancel the
radiation-pressure noise at low frequencies\,\cite{KLMTV2001}. To illustrate how this works, we use the
interferometer for which the input-output relation is given by Equation\,\ref{eq:io_tuned}:
\begin{eqnarray}\nonumber
b_{\zeta}(\Omega)&=&b_1(\Omega)\sin\zeta+b_2(\Omega)\cos\zeta\\\nonumber
&=&e^{2i\phi}[\sin\zeta-{\cal K}(\Omega)\cos\zeta]a_1(\Omega)+
 e^{2i\phi}\cos\zeta \,a_2(\Omega) \\
 & &+e^{i\phi}\cos\zeta\sqrt{2{\cal K}(\Omega)}
 \frac{h(\Omega)}{h_{\rm SQL}}.
\end{eqnarray}
Here the first term, proportional to $a_1$, is the radiation pressure
noise; the second term, proportional to $a_2$, is the shot noise; the
third term is the signal. As we can see, if
the quadrature angle $\zeta$ has the following frequency dependence:
\begin{equation}
\tan\zeta = {\cal K}(\Omega),
\label{eq:bae}
\end{equation}
the radiation pressure term would be canceled, and give rise
to a shot noise only noise floor. Since the phase for the local oscillator
is usually fixed, before beating with the local oscillator we need to rotate
the output quadratures with a filter cavity to achieve such a frequency-dependent
quadrature readout.

\begin{figure}[h]
\centering
    \includegraphics[width=0.6\columnwidth]{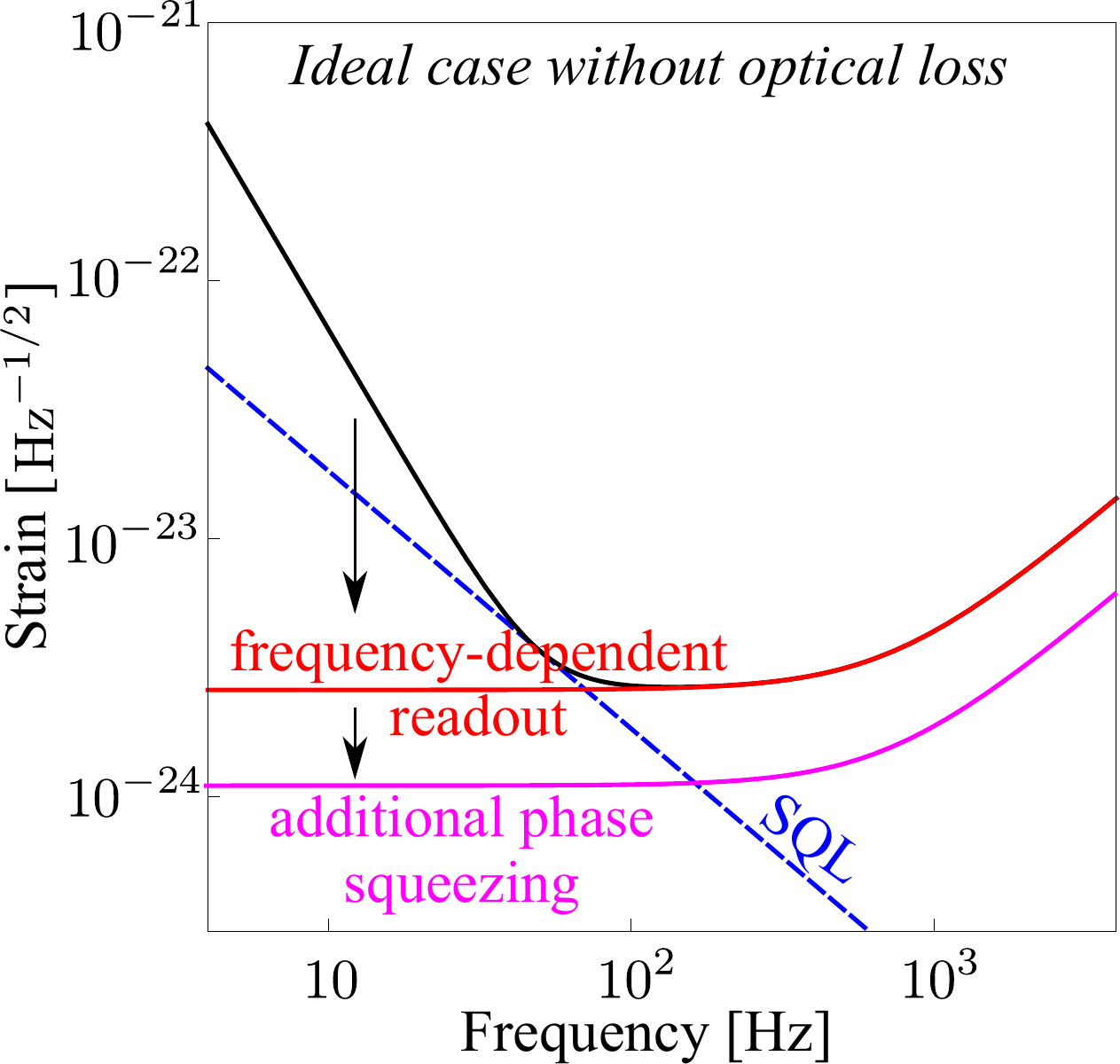}
     \caption{Noise spectra for the frequency dependent readout scheme without (red curve) and
                   with (purple curve) additional phase squeezed light injection.}
\label{fig:Sh_output_filtering}
\end{figure}

The resulting noise spectrum for this scheme is simply:
\begin{equation}
S_h(\Omega)= \frac{1}{{\cal K}(\Omega)}\frac{h_{\rm SQL}^2}{2}.
\label{eq:sh_var}
\end{equation}
If we simultaneously inject phase squeezed light, we will have:
\begin{equation}
S^{\rm opt}_h(\Omega)= \frac{e^{-2r}}{{\cal K}(\Omega)}\frac{h_{\rm SQL}^2}{2}.
\label{eq:sh_var_sqz}
\end{equation}
In Figure~\ref{fig:Sh_output_filtering}, we plot the noise spectrum
in the ideal lossless case with the low frequency radiation pressure noise
completely evaded. In reality, due to optical losses, such a
cancellation cannot be perfect. In the numerical optimization, we will take
into account the optical loss and optimize the parameters for the filter cavity.

\subsubsection{The Effects of Optical Scatter Loss}
\begin{figure}[ht]
  \centering
  \includegraphics[width=\columnwidth]{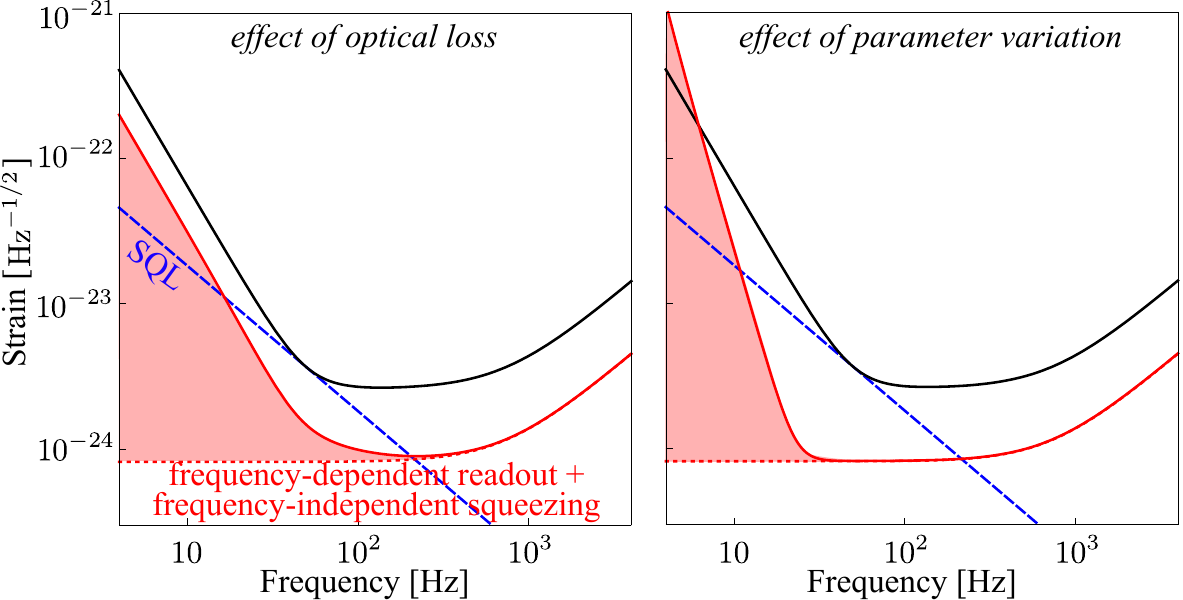}
  \caption{The effect of optical loss in the filter cavity (left) and the effect of parameter
    variation of the filter cavity (right) in the case of frequency-dependent readout
    (output filtering). Similar to the case shown in
    Figure\,\ref{fig:input_filtering_imperfection},
    the shaded areas denote the degradation of sensitivity. We have used a total optical
    loss of ${\cal E}=20\%$. In contrast, the parameter variation is chosen to be only
    ${\delta T_f/T_f^{\rm opt}}={\delta \Delta_f/\Delta_f^{\rm opt}}=10^{-4}$ in order
    to produce reasonable sensitivity, as it is much more sensitive than input filtering.}
  \label{fig:output_filtering_imperfection}
\end{figure}

For the frequency dependent readout scheme, the additional noise introduced
by optical loss influences the output and modifies the input-output relation
in the following way:
\begin{eqnarray}\nonumber
\left[\begin{array}{c}b'_1\\
b'_2\end{array}\right] = \sqrt{{\cal E}}\left[\begin{array}{c}n_1\\n_2\end{array}\right]&+&
\sqrt{1-{\cal E}}\,e^{2i\phi}\left[\begin{array}{cc}1&0\\-{\cal K}
&1\end{array}\right]\left[\begin{array}{c}a_1\\a_2\end{array}\right]
\\&+&\sqrt{1-{\cal E}}\,e^{i\phi}\left[\begin{array}{c}0\\\sqrt{2{\cal K}}\end{array}\right]
\frac{h}{h_{\rm SQL}}.
\end{eqnarray}
Due to the presence of uncorrelated noise, the condition in Equation\,\ref{eq:bae}
no longer provides radiation-pressure noise cancellation. By optimizing the quadrature
angle $\zeta$ for the tuned interferometer with phase squeezed light injection, one can
find that the optimal sensitivity, in contrast to Equation\,\ref{eq:sh_var_sqz}, reads:
\begin{equation}
S_h(\Omega)=\left[\frac{{\eta}\, e^{2r} {\cal K}(\Omega)}{{\eta}+e^{2r}}+
\frac{e^{-2r}+{\eta}}{{\cal K}(\Omega)}\right]\frac{h_{\rm SQL}^2}{2}
\end{equation}
with ${\eta}\equiv {\cal E}/(1-{\cal E})\approx {\cal E}$. The effect of loss is
illustrated in Figure\,\ref{fig:output_filtering_imperfection}. As we can see, the
low-frequency performance is very fragile, and we end
up with a sensitivity similar to the input filtering case, given the same level of loss.

\subsubsection{The effect of parameter variation in the filter cavity}
As shown in \ref{s:tolerance}, the parameter variation of the filter cavity results in the following
sensitivity
\begin{equation}
S_h\approx S_h^{\rm opt}+(2e^{4r}{\cal K}+e^{8r}{\cal K}^3)\left[\left(\frac{\delta T_f}{T_f^{\rm opt}}\right)^2+\left(\frac{\delta \Delta_f}{\Delta_f^{\rm opt}}\right)^2\right]h^2_{\rm SQL}\,.
\label{eq:vsh_out}
\end{equation}
Since ${\cal K}\gg 1$, by comparing Equation\,\ref{eq:vsh_in} with Equation\,\ref{eq:vsh_out},
we can see that the output filtering is
more susceptible to parameter variation than input filtering, which is illustrated in
Figure\,\ref{fig:output_filtering_imperfection}.
%We have had to assume a much smaller parameter variation in order to produce reasonable sensitivity.

\subsection{Long Signal-Recycling Cavity}

\begin{figure}[h]
  \centering
     \includegraphics[width=\columnwidth]{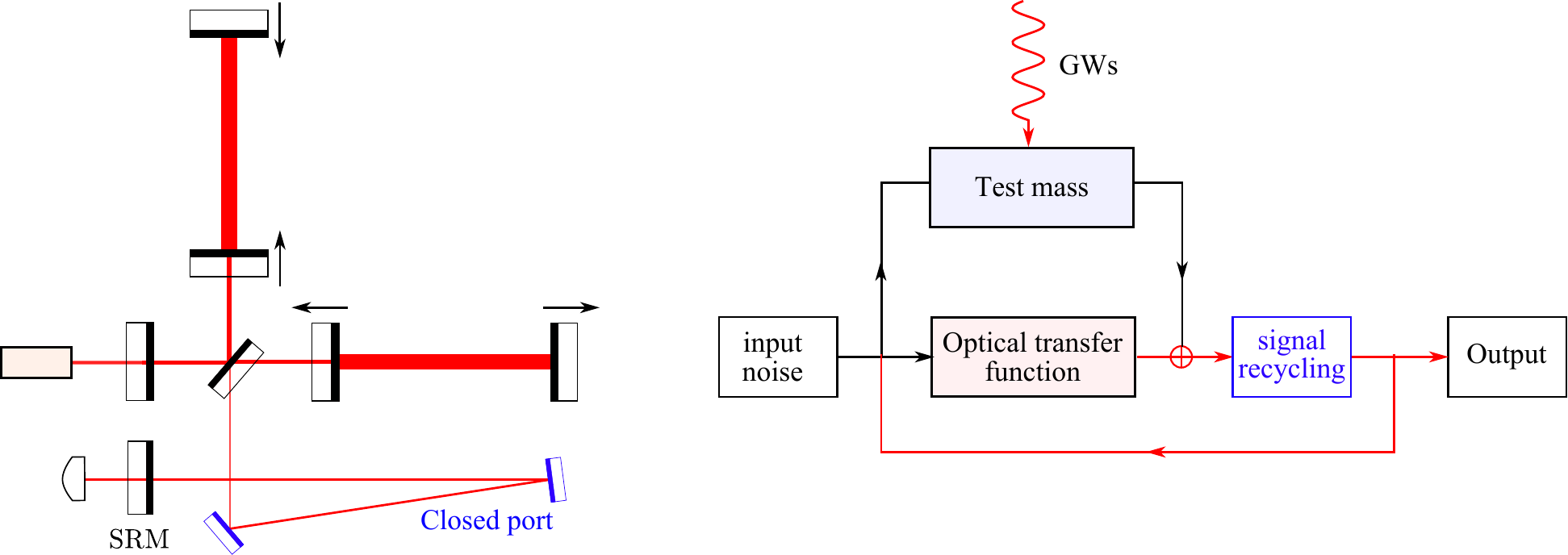}
     \caption{Schematic optical layout showing the long signal recycling cavity scheme (left) and its
       associated block diagram (right). The signal recycling mirror coherently reflects back
       the signal, forming a feedback loop as indicated in the block diagram.}
     \label{fig:config_LSR}
\end{figure}

In this subsection, we discuss the idea of long signal-recycling cavity, and one can refer to Ref.\,\cite{McClelland1995} and references therein for an overview of different recycling techniques applied in the context of GW detectors.
In the usual case when the beam splitter and the signal recycling mirror are close
to each other, the signal recycling cavity is relatively short (order of 10 meters) and
one can ignore the phase accumulated in this cavity by the audio sidebands:
$\Omega L_{\rm sr}/c\approx 0$ with $L_{\rm sr}$ being the length of the signal-recycling cavity.
We can therefore treat the signal-recycling cavity as an effective compound mirror with complex
transmissivity and reflectivity, which is the approach applied in\,\cite{scaling_law}.
With a long signal recycling cavity, however, $\Omega L_{\rm sr}/c$ is not negligible and
different sidebands pick up different phase shifts. Specifically, the transfer function
matrix for the quadratures due to the free propagation in the signal recycling cavity is given by:
\begin{equation}
e^{i\Omega \tau_{\rm sr}}\left[\begin{array}{cc}\cos\Delta \tau_{\rm sr} &  -\sin\Delta \tau_{\rm sr}\\
\sin\Delta \tau_{\rm sr} &\cos\Delta \tau_{\rm sr} \end{array}\right]
\end{equation}
with $\tau_{\rm sr}\equiv L_{\rm sr}/c$ and $\Delta$ being the detuning frequency of the signal recycling
cavity. One can then apply the standard procedure to derive the
input-output relation for this scheme. In general, the final expression is quite lengthy and not illuminating,
so we will not show it here, and will evaluate its noise spectrum numerically. There is one interesting
special case which allows an intuitive understanding. It is when the signal-recycling detuning phase is
equal to $\pi/2$. In this case, the coupled cavity, formed by the signal-recycling cavity and the arm
cavity , has two resonant frequencies located symmetrically around the carrier frequency with their
frequency separation determined by the ITM transmissivity. This case has two advantages: firstly, it
allows an equal and balanced enhancement of both the upper and lower sidebands, in contrast to the
conventional signal-recycling interferometer; secondly, there is no optical-spring effect, as the
signal-recycling cavity is tuned, and the test-mass dynamics is therefore not modified. Such a
case is exactly equivalent to the twin signal-recycling scheme theoretically studied by
Th\"{u}ring {\it et al.}\,\cite{Thuring2007} and experimentally demonstrated by Gr\"{a}f
{\it et al.}\,\cite{Graf2013}, which are motivated by the above mentioned two advantages.

\subsection{Speed Meter with Sloshing Cavity}
\label{s:speedmeter}
The motivation for the
speed meter originates from the perspective of viewing the gravitational-wave detector
as a quantum measurement device. Normally, we measure the test mass position
at different times to infer the gravitational-wave signal. However, position is not
a conserved dynamical quantity of a free mass. According to quantum measurement theory\,\cite{BrKh1999a},
such a measurement process will inevitably introduce additional back action on to the test mass.
In the context here, the back action is the radiation-pressure
noise. In order to evade the back action, one needs to measure the conserved dynamical
quantities of the test mass: the momentum or the energy. Since the momentum is proportional
to the speed, the speed meter can therefore detect gravitational wave without being limited by
the radiation-pressure noise\,\cite{Khalili2}.

\begin{figure}[h]
  \centering
     \includegraphics[width=\columnwidth]{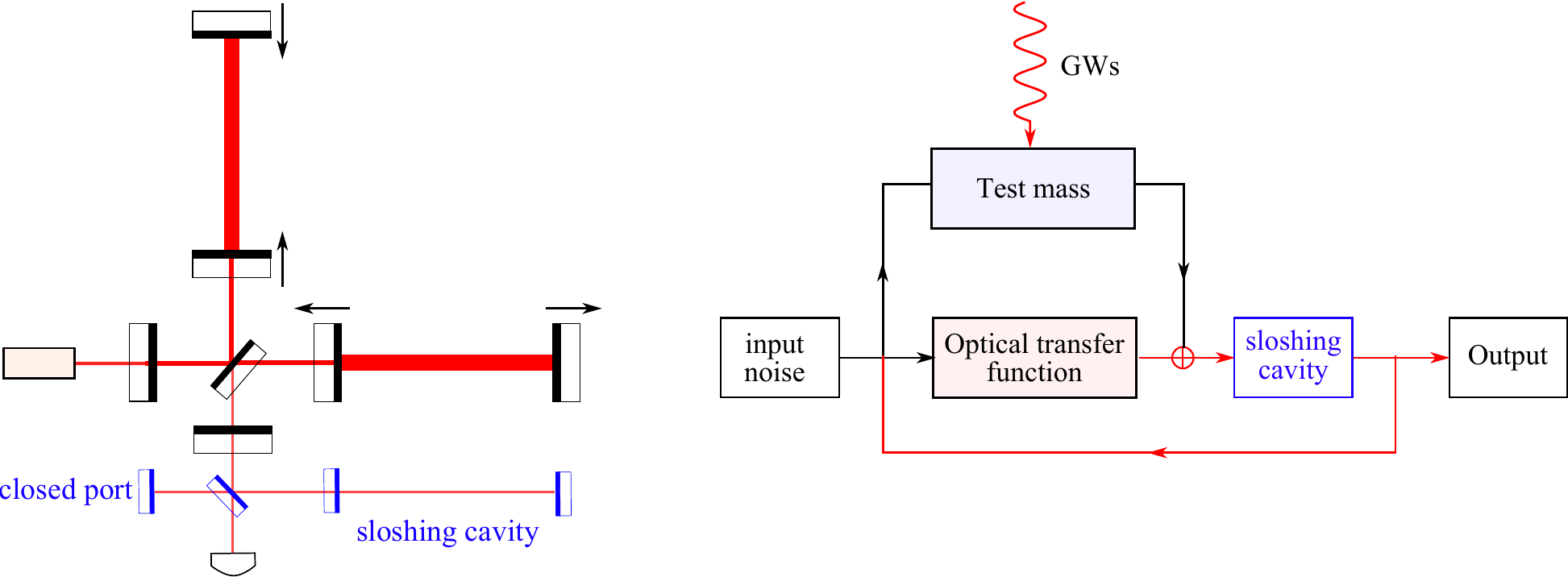}
     \caption{Schematics showing the speed-meter configuration (left) and its block diagram (right).}
  \label{fig:config_speedmeter}
\end{figure}

There are several speed-meter configurations, e.g.,
the Sagnac interferometer\,\cite{Che2003, Beyersdorf1999, Danilishin2004, Chen2011} and a recent
proposed scheme by using different
polarizations\,\cite{Wade2012}. In Figure\,\ref{fig:config_speedmeter}, we show one
particular variant, which is proposed in\,\cite{PuCh2002}, by using a sloshing cavity.
We can gain a qualitative understanding of how such a scheme allows us to measure the speed
of the test mass. Basically, the information of test mass position at an early moment is stored
in the sloshing cavity, and it is coherently superimposed (but with a minus sign due to the phase shift
in the tuned cavity) with the output of the interferometer which contains the current test mass
position. The sloshing happens at a frequency that is comparable to the detection frequency, and
the superposed output is, therefore, equal to the derivative of the test-mass position, i.~e., the speed.

\begin{figure}[h]
  \centering
  \includegraphics[width=0.6\columnwidth]{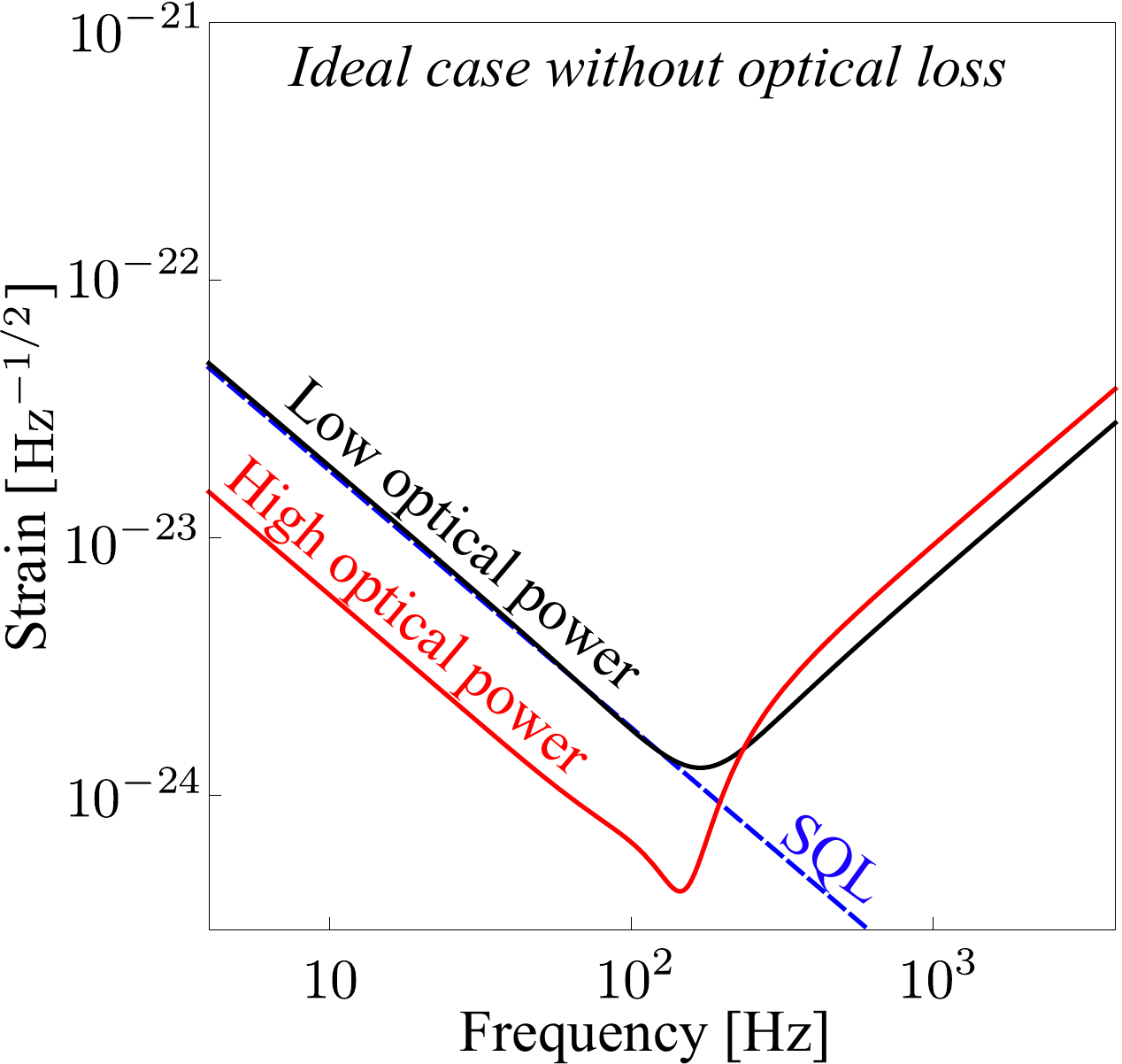}
  \caption{Noise spectrum for the speed-meter configuration with two different levels
                 of circulating optical power.}
    \label{fig:Sh_speedmeter}
\end{figure}

The details of this scheme have been presented in\,\cite{PuCh2002}, in particular the input-output
relation which will be used in the numerical optimization. At this moment, we just show the resulting
quantum-noise spectrum for this scheme:
\begin{equation}
S_h(\Omega)=\frac{(\tan\zeta-{\cal K}_{\rm sm})^2+1}{2{\cal K}_{\rm sm}(\Omega)} {h_{\rm SQL}^2}(\Omega)
\end{equation}
with
\begin{equation}
{\cal K}_{\rm sm}(\Omega)=\frac{16\omega_0\gamma P_c}{m cL[(\Omega^2-\omega_s^2)^2+\gamma^2\Omega^2]}.
\end{equation}
Because ${\cal K}_{\rm sm}$ has a flat frequency response, by properly choosing the homodyne angle $\zeta$, we can
remove the low-frequency radiation pressure noise, and the sensitivity is only
limited by the amount of optical power that we have. This noise spectrum is shown in
Figure\,\ref{fig:Sh_speedmeter}. The low-frequency spectrum has the same slope as the standard
quantum limit, which is a unique feature of speed meter. When the optical power is high enough,
we can surpass the standard quantum limit.

One important characteristic frequency for this type of speed meter
is the sloshing frequency $\omega_s$, and it is defined as
\begin{equation}
\omega_s=\frac{c}{2}\sqrt{\frac{T_s}{L\, L_s}},
\end{equation}
where $T_s$ is the power transmissivity for the front mirror of the
sloshing cavity and $L_s$ is the cavity length. To achieve a speed response in the
detection band, this sloshing frequency needs to be around 100~Hz. For a 4\,km arm
cavity---$L=4000$\,m and 100\,m sloshing cavity---$L_s=100$\,m, it requires the
transmittance of the sloshing mirror to be
\begin{equation}
T_s \approx 30\,{\rm ppm}.
\end{equation}
This puts a rather tight constraint on the optical loss of the sloshing cavity. To
release such a constraint on the optical loss, we can use the fact that $\omega_s$
only depends on the ratio between the transmissivity of the sloshing mirror and
the cavity length and we can therefore increase the cavity length.

In addition, it seems that no filter cavity is needed for speed meter configuration, as the radiation pressure
noise at low frequencies is canceled. However, such a cancellation is achieved by choosing
the homodyne detection angle
\begin{equation}
\zeta=\arctan {\cal K}_{\rm sm}|_{\Omega\rightarrow 0}\,,
\end{equation}
and a high optical power means a large ${\cal K}_{\rm sm}$ and therefore $\zeta$ deviates
from $0$ (the phase quadrature), decreasing sensitivity at high frequencies. With
frequency-dependent squeezing, we can reduce ${\cal K}_{\rm sm}$, or equivalently, $\zeta$,
at low frequencies, which allows us to enhance the high-frequency
sensitivity. Similarly, the frequency dependent readout allows us to cancel the low-frequency
radiation pressure noise without sacrificing the high-frequency sensitivity by
rotating the readout angle to the phase quadrature at high
frequencies.

\subsection{Multiple Carrier Fields}
\begin{figure}[h]
\centering
  \includegraphics[width=\columnwidth]{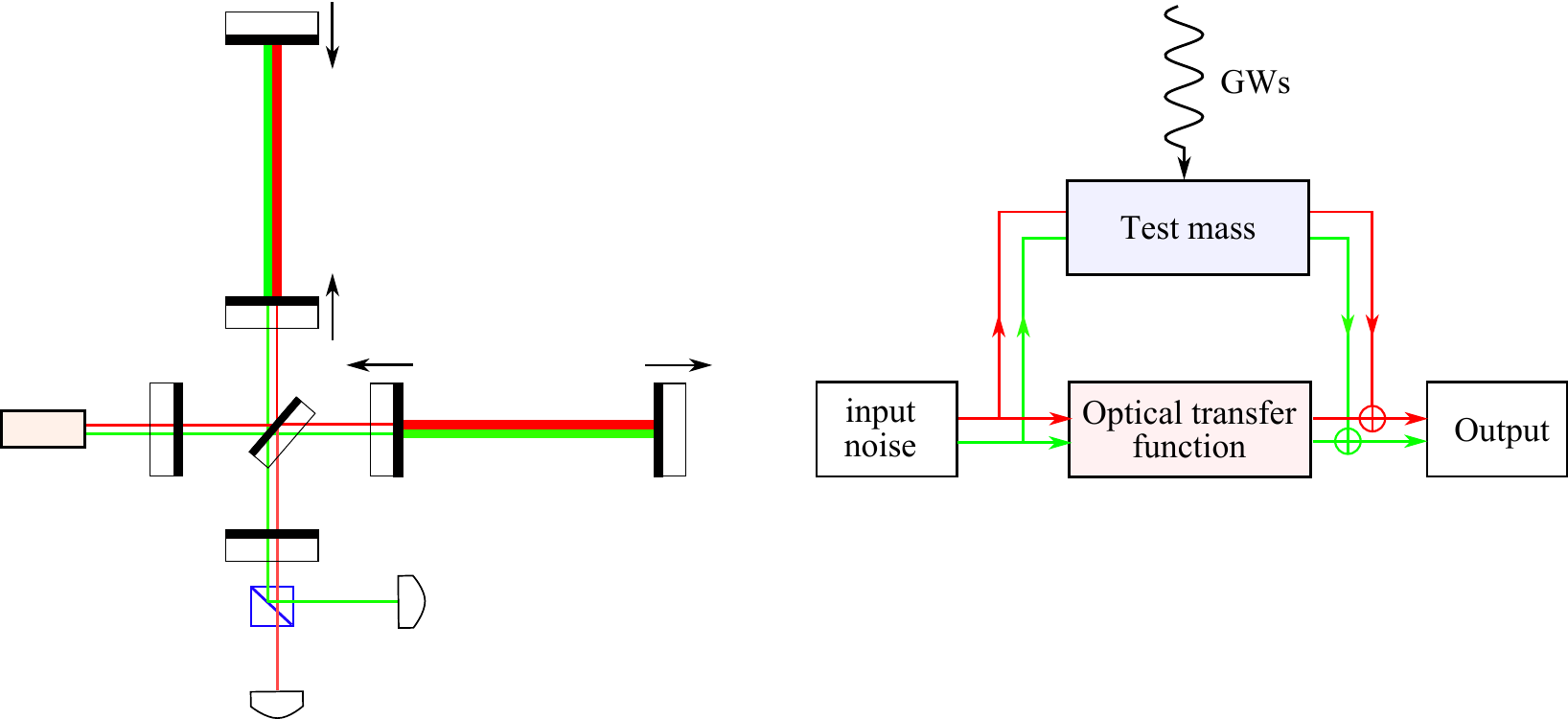}
  \caption{Schematics showing the dual-carrier scheme (left) and its block diagram (right).}
  \label{fig:config_dual_carrier}
\end{figure}

In this section, we will introduce the multiple carrier scheme, and in particular,
we will focus on the dual-carrier case as shown schematically in
Figure\,\ref{fig:config_dual_carrier}.
The additional carrier field provides us with another readout channel. As these two fields
can have a very large frequency separation, we can, in principle, design the optics in
such a way that they have different optical power and see different detuning and bandwidth.
In addition, they can be independently measured at the output. This allows us to gain a lot
flexibilities and effectively provides multiple interferometers but within the same set of optics.

These two optical fields are not completely independent, and they are coupled to each other
as both act on the test masses and sense the test-mass motion (shown pictorially by the
block diagram in Figure\,\ref{fig:config_dual_carrier}). More explicitly, we can look at the input-output
relation for this scheme in the simple case when both fields are tuned (in the SRC):
\begin{eqnarray}\nonumber
\left[\begin{array}{c}b_1^{(A)}\\b_2^{(A)}\\b_1^{(B)}\\b_2^{(B)}\end{array}\right]
&=&\left[\begin{array}{cccc}
1&0&0&0\\
-{\cal K}_A&1&-\sqrt{{\cal K}_A{\cal K}_B}& 0\\
0&0&1&0\\
-\sqrt{{\cal K}_A{\cal K}_B}&0&-{\cal K}_B&1
\end{array}\right] \left[\begin{array}{c}a_1^{(A)}\\a_2^{(A)}\\a_1^{(B)}\\a_2^{(B)}\end{array}\right]
\\&&+\left[\begin{array}{c}0\\\sqrt{2{\cal K}_A}\\0\\\sqrt{2{\cal K}_B}\end{array}\right]\frac{h}{h_{\rm SQL}},
\label{eq:io_dual}
\end{eqnarray}
where we have ignored the uninteresting phase factor $e^{i\phi}$ and we have introduced
\begin{equation}
{\cal K}_A=\frac{16\, \omega_0^{(A)}\gamma_{A} P^{(A)}_c }{m L c\,\Omega^2(\Omega^2+\gamma_{A}^2)},\quad\;
{\cal K}_B=\frac{16\, \omega_0^{(B)}\gamma_{B} P^{(B)}_c }{m L c\,\Omega^2(\Omega^2+\gamma_{B}^2)}.
\end{equation}
The term $-\sqrt{{\cal K}_A{\cal K}_B}$ in the transfer function matrix indicates the coupling between
these two optical fields, and it comes from the fact that the radiation-pressure noise from the first one
is sensed by the second one and vise versa.

As mentioned earlier, because the frequency separation between these two fields is much lager than the
detection band, they
can be measured independently and give two outputs $b_{\zeta}^{(A)}$ and $b_{\zeta}^{(B)}$:
\begin{equation}
b_{\zeta}^{(A)}=b_1^{(A)}\sin\zeta_A+b_2^{(A)}\cos\zeta_A,\quad b_{\zeta}^{(B)}=b_1^{(B)}\sin\zeta_B+b_2^{(B)}\cos\zeta_B.
\end{equation}
To achieve the optimal sensitivity,
we need to combine them with "optimal" filters $C_A(\Omega)$ and $C_B(\Omega)$, obtaining
\begin{equation}
b_{\zeta}^{\rm tot}(\Omega)= C_A(\Omega) b_{\zeta}^{(A)}(\Omega)+C_B(\Omega) b_{\zeta}^{(B)}(\Omega).
\end{equation}
In Ref.~\cite{Yanbei:Local}, the authors have shown the procedure for obtaining the optimal sensitivity and
the associated optimal filters in the general case with multiple carrier fields. Given
the input-output relation: ${\bm b}={\bf M}{\bm a}+{\bm v}h$---a simplified vector form of
Equation\,\ref{eq:io_dual}, the noise spectrum that gives the optimal sensitivity is:
\begin{equation}
S_h(\Omega)=\left[{\bm v}^{\dag}{\bf M}_{\rm hd}^{\dag}({\bf M}_{\rm hd}{\bf M}\,{\bf M}^{\dag}{\bf M}_{\rm hd}^{\dag})^{-1}{\bf M}_{\rm hd}{\bm v}\right]^{-1},
\end{equation}
where we have defined:
\begin{equation}
{\bm M}_{\rm hd}=\left[\begin{array}{cccc}\sin\zeta_A&\cos\zeta_A&0&0
\\0&0&\sin\zeta_B&\cos\zeta_B\end{array}\right].
\end{equation}
This result is used for our numerical optimization in Section~\ref{s:Optimize}.

\subsection{Local Readout}

\begin{figure}[h]
\centering
   \includegraphics[width=\columnwidth]{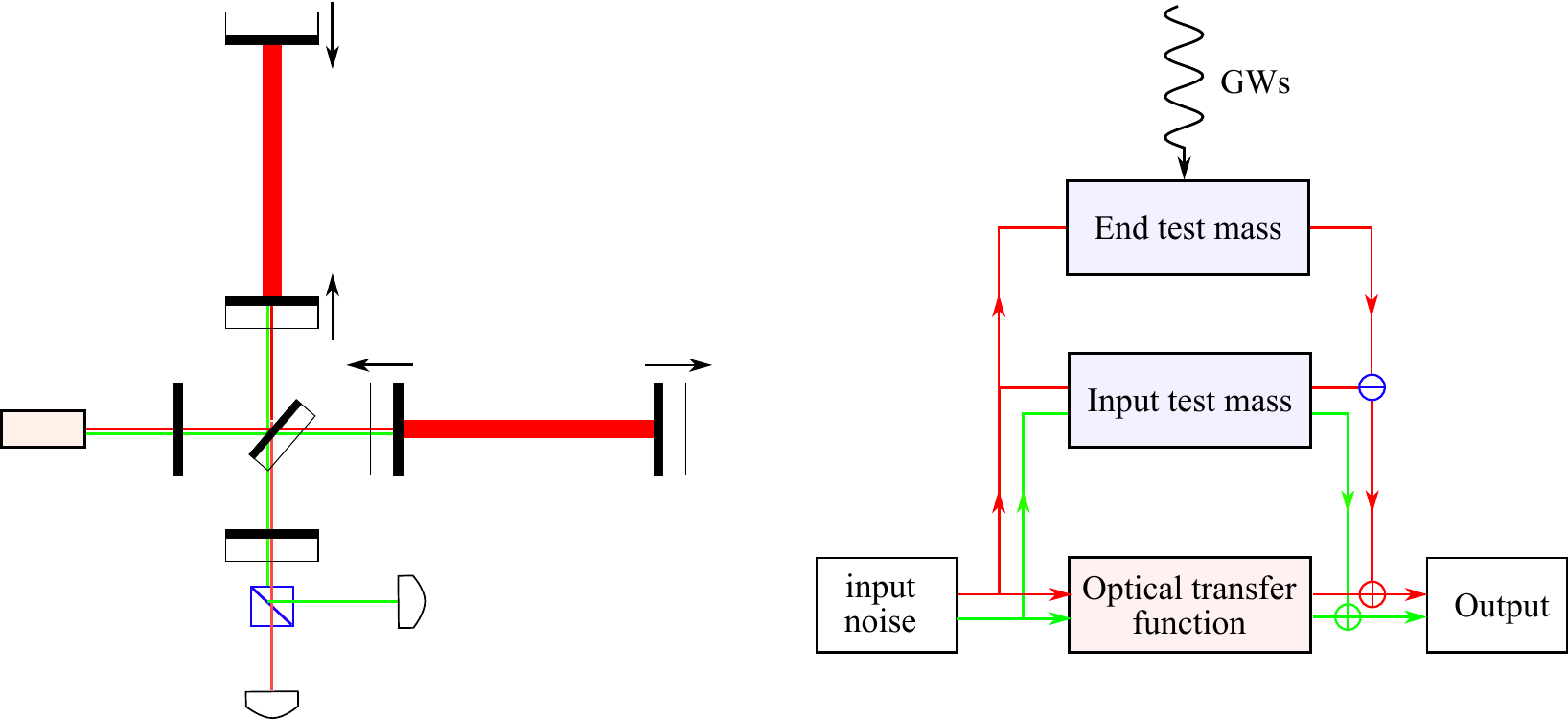}
   \caption{Schematics showing the local-readout topology (left) and the corresponding block diagram (right).}
   \label{fig:config_local_readout}
\end{figure}

Here we will discuss a special case of the multiple-carrier scheme---the local-readout scheme, as shown schematically in
Figure\,\ref{fig:config_local_readout}. In this scheme, the second carrier field is only resonant in the
power-recycling cavity and is
anti-resonant in the arm cavity (barely enters the arm cavity). Why we single this scheme out
of the general dual-carrier scheme and give it a special name is more or less due to a historic reason. This
scheme was first proposed in\,\cite{Yanbei:Local} and was motivated by trying to enhance the
low-frequency sensitivity of a detuned signal-recycling interferometer, which is not as good as the
tuned signal-recycling due to the optical-spring effect. The name ``local readout" originates from
the fact that the second carrier field only measures the motion of the input test mass (ITM) which is {\it local}
in the proper frame of the beam splitter and does not contain the gravitational-wave signal.
One might ask: ``how can we recover the detector sensitivity if the second carrier measures something that
does not contain the signal?" Interestingly, even though no signal is measured by the second carrier, it measures
the radiation-pressure noise of ITM introduced by the first carrier which has a much higher optical power due to
the amplification of the arm cavity, as shown schematically by the block diagram of Figure\,\ref{fig:config_local_readout}.
By combining the outputs of two carriers optimally, we can cancel some part of the radiation-pressure noise and
enhance the sensitivity---the local-readout scheme can therefore be viewed as a noise-cancellation scheme.
The cancellation efficiency is only limited by the radiation-pressure noise of the second carrier field.

To evaluate the sensitivity for this scheme rigorously, one has to treat the input
test mass (ITM) and end test mass (ETM) individually, instead of combining them into an single effective mass
as we did for those schemes mentioned earlier. One can read~\cite{Yanbei:Local} for details.

\newpage
\section{Numerical Optimization}
\label{s:Optimize}
To arrive at the optimum sensitivity (as defined by Equation~\ref{eq:cost_function}, our cost function), 
we use a simplex numerical optimization to vary the optical parameters for 
each of the previously described topologies. For this optimization, we also take into account the 
various classical noise sources (to be distinguished from the quantum noise). 
These noise sources are described in~\ref{s:classical}.

\begin{figure}[h]
\centering
   \includegraphics[width=\columnwidth]{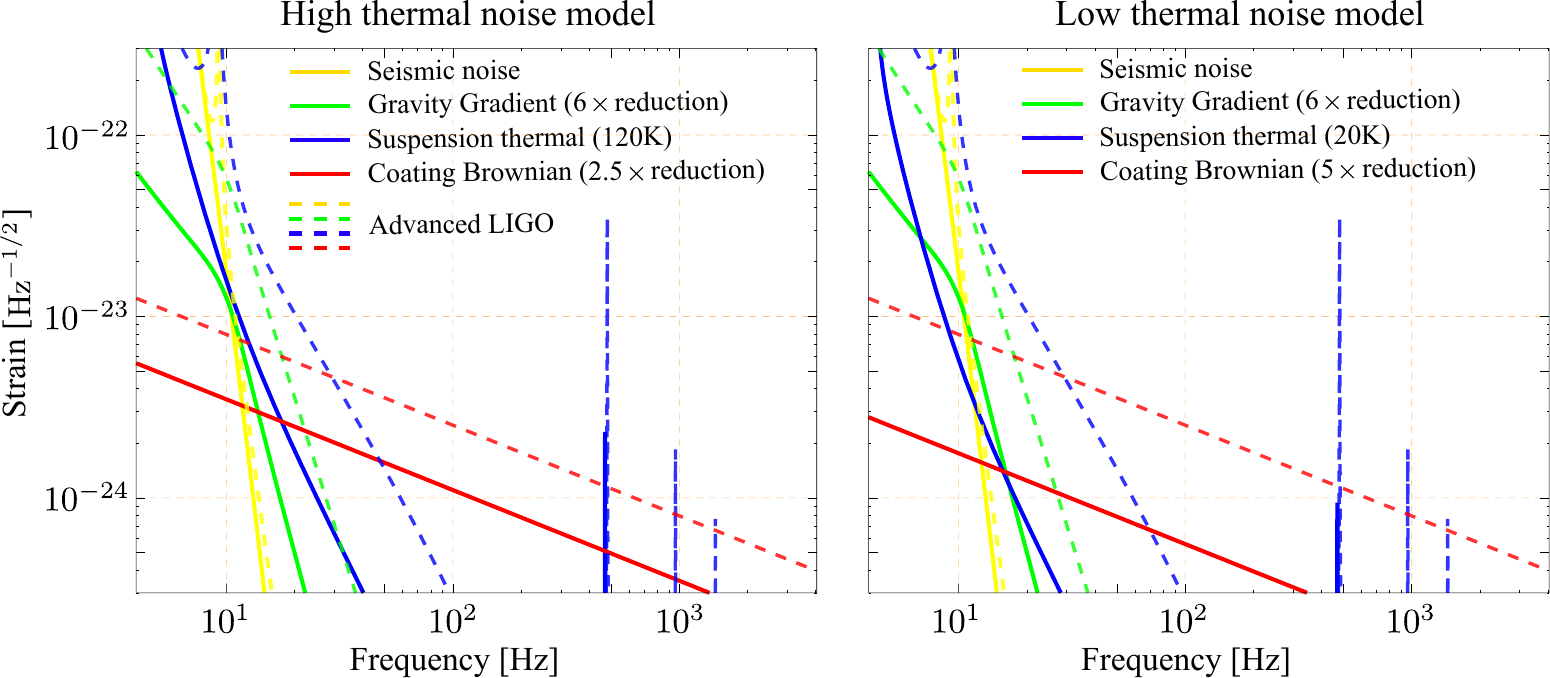}
  \caption{Spectra of  the high classical noise model (left) and the low classical noise model (right).}
    \label{fig:classical_noise}
\end{figure}

\subsection{Cost Function}
The final optimization result critically depends on the cost function. In the literature, 
optimizations have been carried out by using a cost function that is source-oriented---trying 
to maximize the signal-to-noise ratio for particular astrophysical sources. Here we apply a 
rather different cost function, as shown in Eq.\,\ref{eq:cost_function}, that tries to 
maximize the {\it broadband} improvement over aLIGO.

\subsection{Optimization results}
For the optimization, we separate the configurations into two groups:
(i) {\it the frequency-dependent squeezing (input filtering) group,} in which we consider
adding input filter cavities to those configurations mentioned in Section\,\ref{s:Configurations};
(ii) {\it the variational-readout (output filtering) group}, in which we consider adding output filter cavities.
Note that for those multiple-carrier schemes, e.g., the local-readout scheme, the number of filter cavities is
equal to the number of carrier fields, and the number of optimization parameters therefore
increases proportionally. In real implementations, we might specifically design one filter cavity that
is able to simultaneously filter several carrier fields with different filtering parameters, and we can
then reduce the number of optics.

\subsubsection{Total noise spectrum}
The optimization result for the {\it high classical noise model} was shown at the very beginning
(cf. Figure\,\ref{fig:opt_result}).
Notice that, in that plot, we did not show the dual-carrier scheme with both carrier fields resonant in the
arm cavities, and only showed the local-readout scheme in which only one carrier is resonant in the arm cavity. This is due
to the interesting fact that when we fix
the total power of the two carriers, the optimal power for one carrier turns out to be zero---this simply
recovers the single-carrier case. Admittedly, this is due to the specific cost function and the thermal-noise
model that we have chosen. In general, it is not clear that this would be optimal.

The optimization result for the {\it low classical noise model} was shown in Figure\,\ref{fig:opt_result2}.
It is clear that the general features are identical to the input-filtering one. The only
prominent difference comes from the low-frequency sensitivities. This is attributable to the
susceptibility to loss of the frequency dependent readout scheme, as mentioned early in~\ref{s:loss}. Again,
we can see that the speed meter and the local-readout scheme both allow significant improvements
at low frequencies.

In \ref{s:params}, we have listed the optimal values for the different parameters.

\subsubsection{Quantum noise contribution}

\begin{figure}[h]
\centering
\includegraphics[width=\columnwidth]{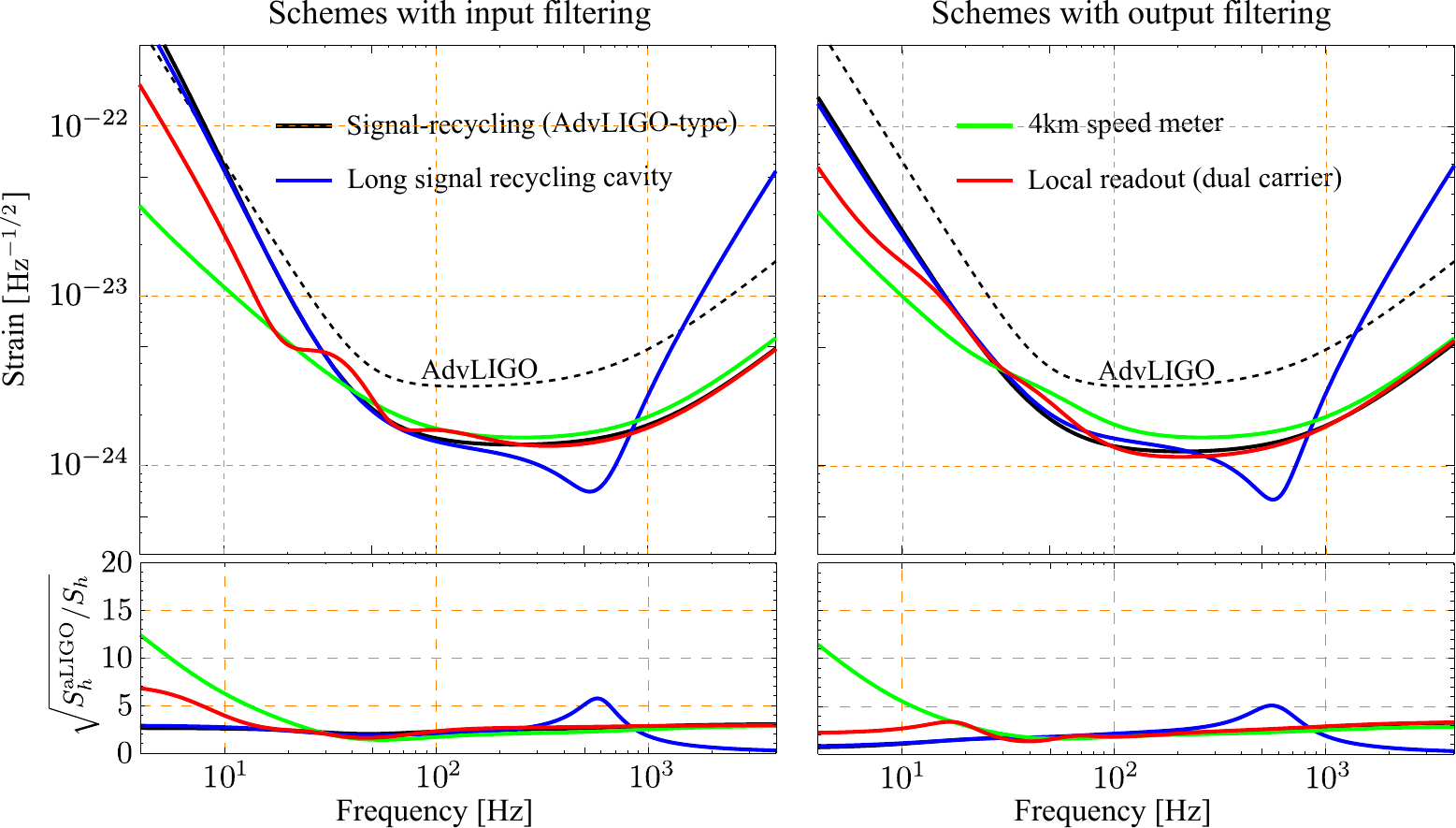}
   \caption{Plot showing the spectra for the quantum noise that contributes to the total noise shown in
   Figure\,\ref{fig:opt_result}.}
   \label{fig:opt_result1_QN}
\end{figure}

\begin{figure}[h]
\centering
\includegraphics[width=\columnwidth]{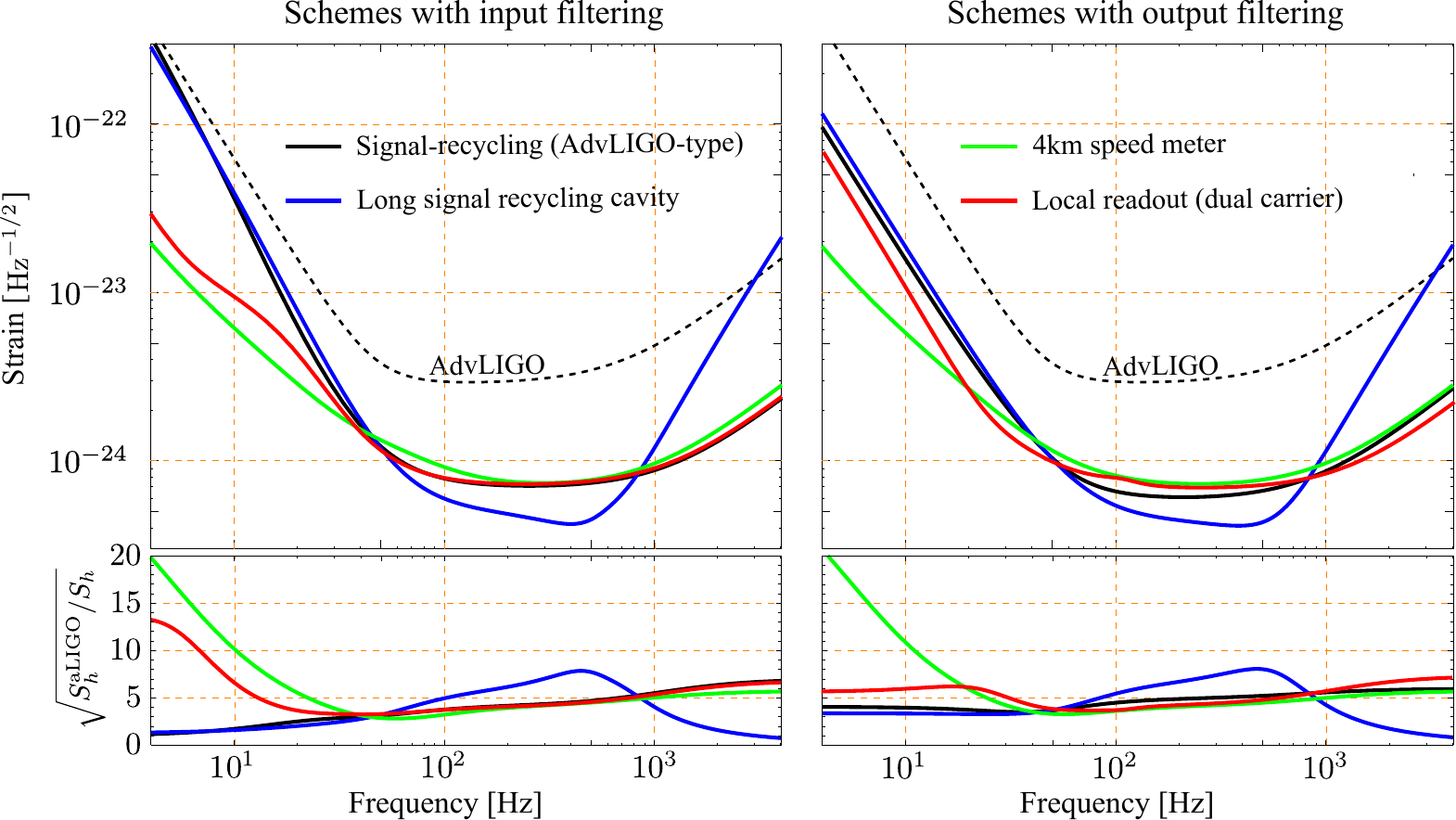}
   \caption{Spectra for the quantum noise shown in Figure\,\ref{fig:opt_result2}.}
   \label{fig:opt_result2_QN}
\end{figure}

To compare the quantum noise contribution to the total noise spectrum, we show only the quantum
noise spectrum in Figure\,\ref{fig:opt_result1_QN} and Figure\,\ref{fig:opt_result2_QN}. It is clear that only
at low frequencies do these schemes differ from each other distinctively. The low-frequency
classical noise masks any difference. Therefore, unless significant
changes can be made to reduce the low-frequency thermal noise, a sound reasoning---for choosing
one advanced configuration over the other as a candidate for upgrade---should be based on the
additional complexity involved, as different schemes do not perform drastically different after taking
into account the thermal noise of the suspensions and mirrors.

\newpage
\section{Future Studies}
\label{s:future}
In the current study, we only cover a few topologies among those that have been
proposed in the literature. To proceed, one approach is to further expand the list
of configurations, but this is a rather daunting task given the huge number of
possible combinations. An alternative that we shall apply in the future is viewing
optical and mechanical components as linear filters, and seeking the answer to
the following question: {\em ``What is the optimal filter that we should place in 
between the test mass and the photodetector such that a specific cost function 
is minimized or if we know the signal waveform?"}
Similar techniques are employed in the design of electronic circuits and optimal
search algorithms for finding signals in noisy data. The
only subtlety is that we are dealing with quantum and classical fluctuations---there 
are certain constraints on the filters that one must apply in order to preserve the 
quantum coherence, especially in cases of amplitude filtering.

\begin{figure}
\centering
    \includegraphics[width=\columnwidth]{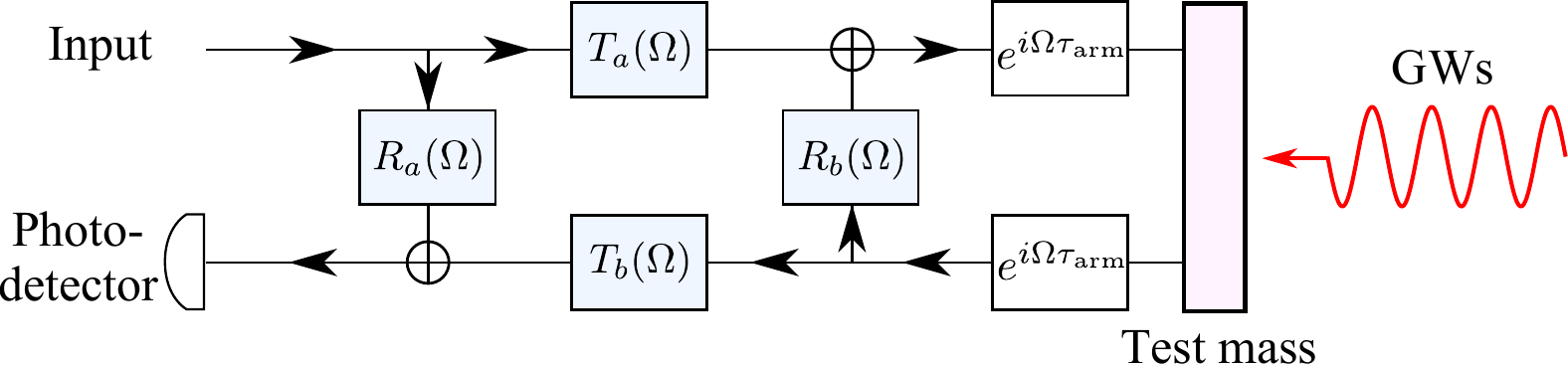}
     \caption{Schematics illustrating the generic four-port filter that can be applied in between the
     test mass and the photodetector. Here we are considering one sideband frequency $\Omega$;
     $\tau_{\rm arm}=L/c$ is the time delay by the interferometer arm.}
    \label{fig:detect}
\end{figure}

To be concrete, let us look at the structure of the detection process more carefully.
The test mass, which contains the GW signal, turns ingoing optical
fields into outgoing fields which in turn are detected by the photodetector. In between the
test mass and the photodetector, the most generic filter we can apply is a four-port
filter, as illustrated in Figure\,\ref{fig:detect}. The transfer functions of such a four-port
filter---$T_a(\Omega)$, $T_b(\Omega)$, $R_a(\Omega)$ and $R_b(\Omega)$\,\footnote{For simplicity,
here we use the sideband picture instead of quadrature, otherwise these transfer functions
will be transfer matrices.}---are not independent and need to satisfy the Stokes relation
due to energy conservation and time-reversal symmetry. Specifically, if we separate their
amplitude and phase as follows:
\begin{eqnarray}\nonumber
T_a(\Omega) &=& |T_a(\Omega)|e^{i\,\phi_a(\Omega)}, \quad T_b(\Omega) = |T_a(\Omega)|e^{i\,\phi_b(\Omega)},\\
\quad R_a(\Omega) &=& |R_a(\Omega)|e^{i\,\varphi_a(\Omega)}, \quad R_b(\Omega) = |R_b(\Omega)|e^{i\,\varphi_b(\Omega)},
\end{eqnarray}
the Stokes relation dictates the following constraints:
\begin{eqnarray}\nonumber
 |T_a(\Omega)| &=&  |T_b(\Omega)|, \quad  |R_a(\Omega)|  =  |R_b(\Omega)|,\quad  |T(\Omega)|^2 +  |R(\Omega)|^2=1, \\
 e^{i\,\phi_a(\Omega)} &=& e^{i\,\phi_b(\Omega)}, \quad  e^{i\,\varphi_a(\Omega)+i\,\varphi_b(\Omega)} = -e^{2i\,\phi_a(\Omega)}.
\end{eqnarray}

\begin{figure}
\centering
    \includegraphics[width=\columnwidth]{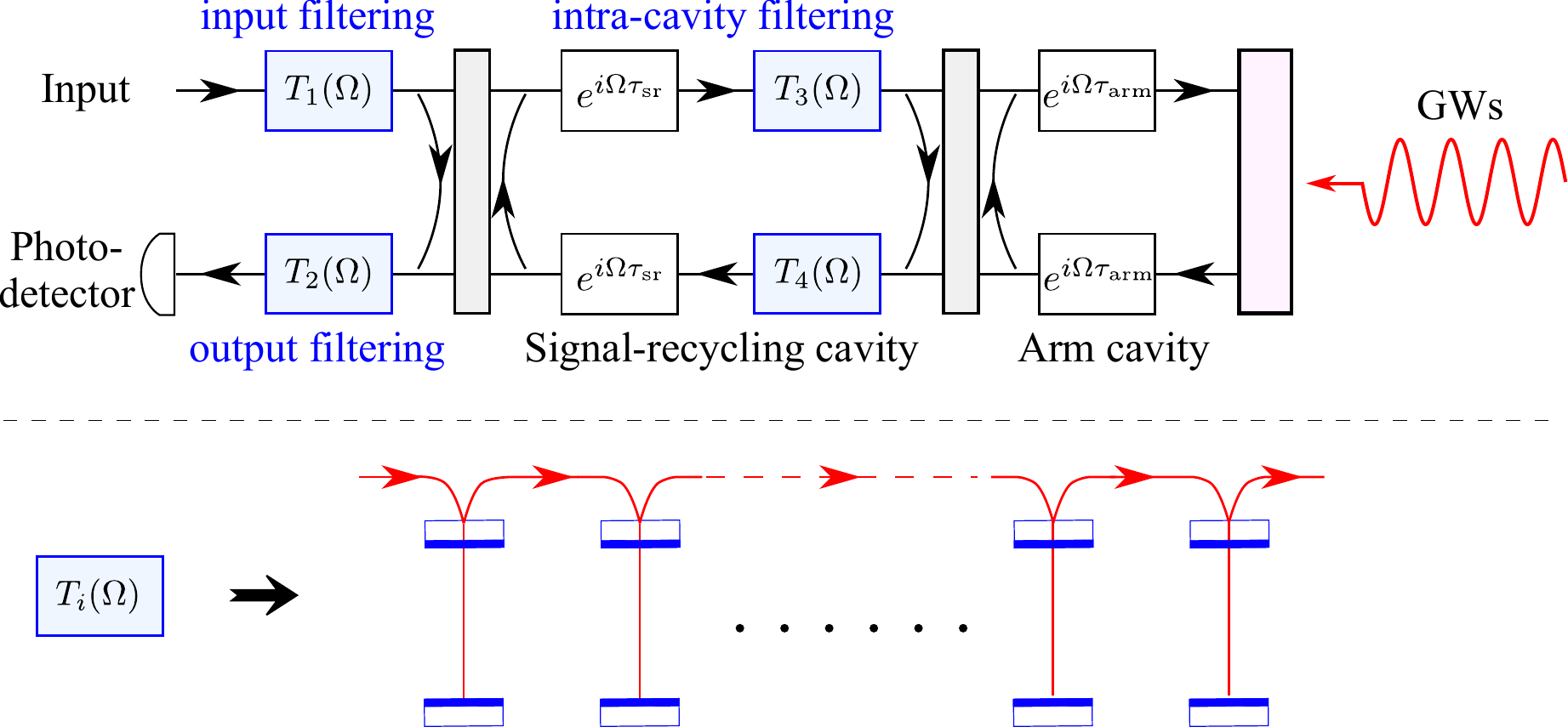}
     \caption{Schematics illustrating the scheme that we will numerically optimize (top). Each
     of these transfer functions corresponds to a cascade of (opto-mechanical) cavities in series (bottom).}
    \label{fig:opt_config}
\end{figure}

In order to obtain the optimal four-port filter given a certain cost function, we can either (i) parameterize
those transfer functions in terms of zeros and poles and optimize them ---
this requires a mapping between zeros and poles to the physical setup, which is highly nontrivial,
or (ii) insert a number of cavities and optimize the parameters --- this is
more transparent in terms of finding out the physical scheme. As a first
attack, we will apply the latter approach, as illustrated in Figure\,\ref{fig:opt_config}.
Not only do we consider input filtering $T_1(\Omega)$ and output filtering $T_2(\Omega)$,
we also include the intra-cavity filtering $T_3(\Omega)$ and $T_4(\Omega)$ --- the filters sit
inside the signal-recycling cavity (the sloshing cavity in the speed-meter configuration is one
special example of the intra-cavity filtering). These filters are different cascades of cavities that can
either have fixed mirrors (the passive cavity) or a movable end mirror (the opto-mechanical cavity).
The usual passive optical cavity only allows us to create a frequency-dependent phase shift on
the sidebands, or equivalently, frequency-dependent rotation of the amplitude and phase
quadratures. By adding control light and allowing the end mirror to be movable, we can also
create frequency-dependent amplitude modulation, similar to the ponderomotive squeezer
proposed in\,\cite{Corbitt2006}. Recently such active cavities with opto-mechanical interactions have triggered
interesting discussion within the GW community, as it allows us to filter the
audio-band signal with table-top scale setups. However, to realize it experimentally,
the mirror thermal noise needs to be low enough such that the quantum coherence shall not be
destroyed. This probably requires cryogenic temperatures which is somewhat challenging to realize.
In future numerical optimization, we will study the influence of thermal noise in the opto-mechanical
cavity on the sensitivity.

\section{Conclusions}
\label{s:Conclusion}
We have optimized the quantum noise spectrum for a few different interferometer configurations that are
candidates for the 3$^{\rm rd}$ generation LIGO. In particular, we have considered the frequency
dependent squeezing (input filtering) and frequency dependent readout (output filtering);
introducing additional filter cavities either at the input or the output ports. Limited
by thermal noise at low frequencies, the difference among these configurations is not very prominent.
This leads us to the conclusion that adding one input filter cavity to Advanced LIGO seems to be the most
feasible approach for upgrading in the near term, due to its simplicity compared with other schemes.
If the low-frequency thermal noise can be reduced in the future, the speed meter and the multiple-carrier
scheme can provide significant low-frequency enhancement of the sensitivity. This extra enhancement
will, for some low enough thermal noise, be enough to compensate for the extra complexity.

\ack
We would like to thank our colleagues in the LIGO Scientific Collaboration's Advanced 
Interferometer Configurations
working group for fruitful discussions.
R.~X.~A. is supported by NSF grant PHY-0757058.
H.~M., H.~Y., and Y.~C. are supported by NSF grants PHY-0555406,
PHY-0653653, PHY-0601459, PHY-0956189, PHY-1068881,
as well as the David and Barbara Groce startup fund at Caltech.

\section*{References}
\bibliography{GWreferences}

% Appendix
\appendix

% parameters of the optimization results
\section{Optimal Parameters}
\label{s:params}
Optimal parameters for the configurations with frequency dependent squeeze angle
in the high thermal noise model are listed in Table~\ref{t:isqz_high}. The nominal parameters
common among different configurations are: $m=50$~kg,  $T_{\rm PRM}=0.03$,  $T_{\rm ITM}=0.01$
and maximal input power is equal to 125~W, which corresponds to around 1~MW circulating
in the arm cavity. For the squeezed light, we have assumed 10~dB squeezing with 5\% injection loss, arising 
from the lossy optics between the squeezer and the input mirror of the filter cavity. For the speed meter, 
the length of the sloshing cavity is 4~km and the power
transmittance for the sloshing mirror is equal to 700~ppm. Note that all the configurations are tuned, as
a broadband sensitivity is preferred, given the particular cost function that we have chosen.
\begin{table}[h]
\centering
\begin{tabular}{|c|c|c|c|c|c|}
\hline
 Configurations & $P_0$ (W) & $T_{\rm sr}$ & $T_{f}$ (ppm) & $\Delta_f$ (Hz) & FOM \\
 \hline
 Signal-recycling & 125.0 & 0.12& 249.2 & $-$28.5 & 206.5 \\
\hline
Long signal recycling  & 125.0 & 0.1 & 179.2 & $-$21.7 & 186.5 \\
\hline
Speed meter & 125.0 & 0.07 & 334.1 & $-$24.2 & 202.4\\
\hline
Local readout (carrier A) &125.0  & 0.14&164.0 & $-$19.0 & 207.2 \\
Local readout (carrier B) & 46.2 & 0.0013 & 145.0 & $-$13.6 & -\\
\hline
\end{tabular}
  \caption{Optimal parameters for different configurations with frequency-dependent squeezing in the high
  thermal noise model. Here $P_0$ is the input optical power, $T_{\rm sr}$ is the power transmittance
  of the signal recycling mirror, $T_f$ is the power transmittance for the front mirror of the filter cavity,
  $\Delta_f$ is the detune frequency of the filter cavity, and figure of merit (FOM) is equal to $10^6/{\cal C}$ with ${\cal C}$
  the value of the cost function defined in Eq.\,\ref{eq:cost_function}---the larger the figure of merit is, the
  better the broadband sensitivity is. }
  \label{t:isqz_high}
\end{table}

The optimal parameters in the high thermal noise model for different
configurations with frequency dependent (variational) readout quadrature are listed in the Table~\ref{t:var_high}.
The common parameters are the same as those for the frequency dependent squeezing.
The transmissivity of the sloshing mirror for the speed meter is equal to 900~ppm.
\begin{table}[h]
\centering
\begin{tabular}{|c|c|c|c|c|c|}
\hline
 Configurations & $P_0$ (W) & $T_{\rm sr}$  & $T_{f}$ (ppm) & $\Delta_f$ (Hz) & FOM \\
 \hline
 Signal-recycling & 125.0 & 0.1  & 233.3 & $-$25.7 & 196.4 \\
\hline
Long signal recycling  & 125.0 & 0.10 & 231.6 & $-$25.5 & 179.3 \\
\hline
Speed meter & 125.0 & 0.27 & 321.7 & $-$16.7 &202.7 \\
\hline
Local readout (carrier A) &125.0  & 0.12 & 858.5 & $-$22.5 & 205.2 \\
Local readout (carrier B) & 38.0 & $0.001$ & 858.1 & $-$21.9 & -\\
\hline
\end{tabular}
 \caption{Optimal parameters for different configurations with frequency dependent readout in
                the high thermal noise model.}
  \label{t:var_high}
\end{table}

The optimal parameters for different configurations with frequency dependent squeezing
in the low thermal noise model are listed in the Table~\ref{t:isqz_low}.
The common parameters for different configurations are: $m=150$~kg, $T_{\rm PRM}=0.03$,
$T_{\rm ITM}=0.01$ and maximal input power is equal to 500~W, which corresponds to
approximately 3~MW intra cavity power. The specification for the squeezing is the same as the 
low-noise model case---10~dB squeezing with 5\% injection loss. The 
optimal sloshing mirror transmittance for the speed meter is equal to 970~ppm.

\begin{table}[h]
\centering
\begin{tabular}{|c|c|c|c|c|c|}
\hline
 Configurations & $P_0$ (W) & $T_{\rm sr}$  & $T_{f}$ (ppm) & $\Delta_f$ (Hz) & FOM \\
 \hline
 Signal-recycling & 500.0 & 0.12 & 256.9 & $-$31.4 & 259.9 \\
\hline
Long signal recycling  & 500.0 & 0.19 & 372.4 & $-$43.6 & 240.8 \\
\hline
Speed meter & 500.0 & 0.28 & 335.4 & $-$20.1 & 260.2\\
\hline
Local readout (carrier A) &500.0  & 0.092 &  234.1 & $-$22.1 & 267.9 \\
Local readout (carrier B) & 494.2 & $0.0001$ & 135.8 & $-$49.0 & -\\
\hline
\end{tabular}
 \caption{Optimal parameters for different configurations with frequency-dependent squeezing 
                in the low thermal noise model.}
  \label{t:isqz_low}
\end{table}

The optimal parameters for the low-noise model for different
configurations with frequency dependent readout quadrature are listed in the Table~\ref{t:var_low}.
The optimal sloshing mirror transmittance for the speed meter is equal to 0.0013.

\begin{table}[h]
\centering
\begin{tabular}{|c|c|c|c|c|c|}
\hline
 Configurations & $P_0$ (W) & $T_{\rm sr}$  & $T_{f}$ (ppm) & $\Delta_f$ (Hz) & FOM \\
\hline
 Signal-recycling & 500.0 & 0.087 & 247.5 & $-$27.4 & 249.3 \\
\hline
Long signal recycling  & 500.0 & 0.16 & 335.9 & $-$37.0 & 228.2 \\
\hline
Speed meter & 500.0 & 0.16 & 516.5 & $-$18.4 & 257.5\\
\hline
Local readout (carrier A) & 424.0  & 0.097 & 245.2 & $-$25.9 & 257.6  \\
Local readout (carrier B) & 16.2 & 0.033 & 1000 & 14.1 & -\\
\hline
\end{tabular}
 \caption{Optimal parameters for different configurations with frequency dependent readout
                quadrature in the low thermal noise model.}
  \label{t:var_low}
\end{table}

% parameters for the classical noise sources
\section{Other Noise Sources}
\label{s:classical}

In addition to the quantum noise arising from the fluctuations in the ground state of the electromagnetic field,
the sensitivity of the interferometers is also limited by Brownian thermal noise in the mirror
suspensions~\cite{SUS:2002, SUS:2012},
seismic vibrations propagating to the mirror~\cite{RobEA2004},
terrestrial gravitational fluctuations~\cite{Saulson:NN, NN:subtract2012}, and Brownian thermal
fluctuations of the mirror (and mirror coating) surface~\cite{Harry:CQG2007,Ting:Brownian}. 
In Table\,\ref{t:classical}, we show the physical interferometer parameters used in the optimization.
\begin{table}[h]
\centering
\begin{tabularx}{\columnwidth}{|X|X|X|X|X|X|X|X|}
\hline
Config  &  $M$ (kg) & $T_{\rm mir}$ (K)   & $T_{\rm sus}$ (K)  & $w_{\rm beam}$ (cm) & $\phi_{\rm coat}^{\rm high} \times 10^{-5}$ &$\phi_{\rm coat}^{\rm low} \times 10^{-5}$ & $NN_{\rm FF}$\\
\hline
aLIGO                           &   40   & 295   &   295   & 5.9 &$23$ & $4.0$  & --\\
\hline
High-noise model\,\ref{fig:opt_result}  &   50   & 295   &   120   & 14.6 &$23$ & $4.0$ & 6\\
\hline
Low-noise model\,\ref{fig:opt_result2}  & 150    & 120   &    20   & 5.9 &$2.0$ & $2.0$ & 6\\
\hline
\end{tabularx}
\caption{Physical parameters used for computing the non-quantum noise sources during the configuration optimization. 
              Here $M$ is the mass of a single test mass;  $T_{\rm mir}$ and $T_{\rm sus}$ are the temperature of the mirrors 
              (test masses) and the suspension wire, respectively; $w_{\rm beam}$ is the average beam spot radius on
              the arm cavity mirrors;
              $\phi^{\rm high}_{\rm coat}$ and $\phi^{\rm low}_{\rm coat}$ are the 
              mechanical loss angle for the high (low) refractive index materials, respectively; $NN_{\rm FF}$ 
              is the Newtonian noise subtraction factor achieved by feed-forward cancellation.}
\label{t:classical}
\end{table}

% some discussion about the loss effect
\section{Optical loss and optimal filter cavity length}
\label{s:loss}
It is essential to gain a full understanding---through experiments and
numerical modeling---of how the optical loss scales with the cavity
length, and this will determine the cavity length for achieving the optimal
sensitivity. Here we will provide a qualitative estimate of the dependence
of sensitivity on the optical loss and the cavity length, the connection between
which is left for future work.

\subsection{Qualitative picture}
Given a filter cavity, the detector sensitivity,
is affected by the total loss of the cavity $\cal E$ which is equal to the round-trip loss
$\epsilon$ multiplied by the number of round trips $N\sim 1/T_f$ with $T_f$
being the transmittance of the cavity input mirror (assuming a totally reflected end mirror),
namely
\begin{equation}\label{eq:loss_E}
{\cal E } \approx \frac{\epsilon}{T_f}.
\end{equation}

In addition, since the filter cavity bandwidth $\gamma_f$ needs to be comparable to the detection
bandwidth $\gamma$ of the interferometer in order to reduce the quantum noise, we require $
\gamma_f ={c T_0}/({4L_f})\approx \gamma, $ where $L_f$ is the filter cavity length. It follows that
\begin{equation}
T_f \approx \frac{4 \gamma L_f}{c}.
\end{equation}
Therefore, the total optical loss is given by:
\begin{equation}\label{eq:total_loss}
{\cal E} \approx \frac{c\,\epsilon}{4\gamma L_f}\propto\frac{\epsilon}{L_f},
\end{equation}
which means that the total optical loss depends on the ratio between the round-trip loss and the filter
cavity length.

\begin{figure}[h]
\centering
   \includegraphics[width=\columnwidth]{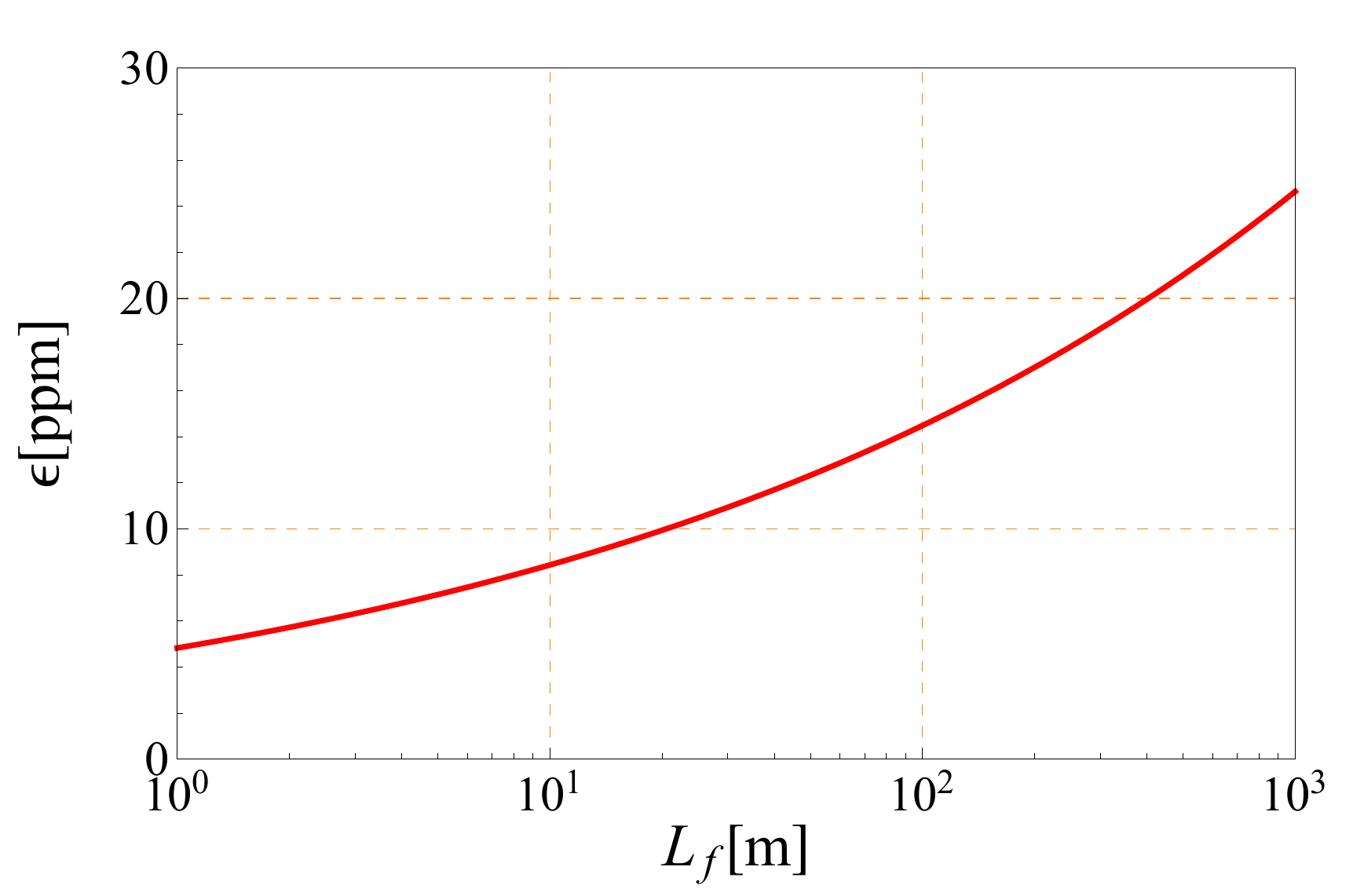}
       \caption{Dependence of scattering loss on the length of the filter cavity due to
                     micro-roughness of the mirror surface. Figure errors at large spatial frequencies are neglected here.}
   \label{fig:length_dependence_loss}
\end{figure}

If the optical loss were independent of the cavity length, the above scaling would imply that the longer the cavity,
the smaller the total loss. However, this is usually not the case. Due to the roughness of the mirror surface, there will
be an optical loss associated with scattering off the surface. One signifcant contribution comes from
the ``figure error'' which
corresponds to roughness with a spatial scale larger than a few millimeters; another is from micro-roughness due to
small-scale imperfections which induce large-angle scattering. This scattering critically depends on the beam size
which in turn depends on the cavity length~\cite{Evans:Losses}. To be more specific, the beam size $w_f$ for a confocal
cavity\footnote{A confocal cavity allows the minimal beam size on
the mirror (see e.g. Chapter 19 in \cite{Siegman:Lasers}).} is given by:
\begin{equation}
w_f=\left(\frac{L_f\lambda}{2\pi}\right)^{1/2}\,.
\end{equation}
The dependence of scattering loss on the beam size is more involved. Here we only discuss the scattering associated
with micro-roughness (the one from ``figure error” requires a numerical simulation and normally does not have a compact
analytical expression).The corresponding scattering loss for an isotropic surface topology reads\,\cite{Yamamoto:07}:
\begin{equation}
\epsilon = \left(\frac{4\pi}{\lambda}\right)^2 \int^{f_s^{\rm max}}_{f_s^{\rm min}}S_1(f_s)\,{\rm d}f_s.
\end{equation}
Here $f_s$ has the units of $m^{-1}$ and is the spatial frequency; $f_s^{\rm min}$ is the minimum spatial frequency, or equivalently, the
maximum length scale of the beam, and it is inversely proportional to the beam size:
\begin{equation}
f_s^{\rm min}\approx (w_f/2)^{-1}
\end{equation}
and the maximum spatial frequency $f_s^{\rm max}$ is taken to be $1/\lambda$.
The one dimensional scattering power spectral density $S_1(f_s)$ depends on specific optics that we use. For initial
LIGO mirrors, the measured power spectral density approximately satisfies\,\cite{Walsh:99}:
\begin{equation}
S_1(f_s)=7\times 10^{-19}f_s^{-1.45}\,.
\end{equation}
In Figure\,\ref{fig:length_dependence_loss}, we show how the optical loss depends on the filter cavity length (numerically
integrating the above integral). It can be approximated by the following power law:
\begin{equation}\label{eq:loss_approximation_formula}
\epsilon|_{\rm micro-roughness} \approx 5.0\, L_f^{0.23}\,.
\end{equation}
In this case, the total optical loss $\cal E$ [cf. Equation\,\ref{eq:total_loss}] scales as $L_{f}^{-0.77}$. This seems to
imply the longer the cavity the better. However, this ignores scattering loss from figure errors and also diffraction.
In reality,  to determine the optimal cavity length, we need to experimentally investigate the dependence of the
round-trip optical loss on the cavity length.

\subsection{Detailed analysis}
\begin{figure}[h]
\centering
   \includegraphics[width=0.5\columnwidth]{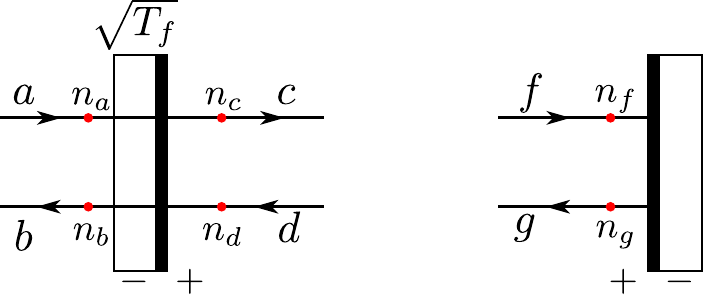}
       \caption{Figure illustrating the propagating fields in a single lossy filter cavity. The red dots are the lossy
                      ports where vacuum fluctuations $n_{\alpha_s}$ that are uncorrelated with the main fields
                      $\alpha_s$ enter ($\alpha_s=a, b, c, d, f, g$) the optical paths.}
   \label{fig:loss}
\end{figure}

Here we can provide a more detailed analysis to elaborate on the qualitative picture that we showed in the
previous section concerning the magnitude of the loss. To account for the optical loss of filter cavities
carefully, we need to inject vacuum fluctuations at every port of all optics\,\footnote{Note that
in the literature, normally one introduces a so-called lossy mirror  to account for all the optical loss and assumes
other mirrors are lossless. This works generally, but may fail when the optical path is complicated. Also there is
some ambiguity in determining the loss for such a effective mirror.}, summarizing the effect from
scattering and absorption. The simple linear lossy cavity is illustrated in Figure\,\ref{fig:loss}, and we can
derive its input-output relation from which we can determine quantitatively how the loss influences the quantum
coherence of the squeezed light (for input filtering) or the output quadratures (for output filtering).

To derive the input-output relation, we use the continuity condition of the fields, which goes as follows:
\begin{eqnarray}\nonumber
\bm b&=& (1-{\epsilon}/{4})(-\sqrt{R_f} \bm a+\sqrt{T_f}\bm d)+\sqrt{({\epsilon}/{4})(1-{\epsilon}/{4})}(-\sqrt{R_f}\,\bm n_a \\ &&+\sqrt{T_f}\,\bm n_d) +\sqrt{{\epsilon}/{4}}\,\bm n_b\,, \\\nonumber
\bm c&=& (1-{\epsilon}/{4})(\sqrt{R_f} \bm d+\sqrt{T_f}\bm a)+\sqrt{({\epsilon}/{4})(1-{\epsilon}/{4})}(\sqrt{R_f}\,\bm n_d \\&&+\sqrt{T_f}\,\bm n_a) +\sqrt{{\epsilon}/{4}}\,\bm n_c\,,\\
\bm f&=& {\bf M}_{\rm rot} \,\bm c\,, \\
\bm g&=& (1-{\epsilon}/{4}) \bm f+\sqrt{({\epsilon}/{4})(1-{\epsilon}/{4})}\, \bm n_f + \sqrt{{\epsilon}/{4}}\,\bm n_g\,,\\
\bm d &=& {\bf M}_{\rm rot}\,\bm g\,.
\end{eqnarray}
Here $\bm \alpha = (\alpha_1,\; \alpha_2)^{\rm T}$ is the vector for the amplitude and phase quadratures;
we have assumed all the ports have identical loss equal to $\epsilon/4$ (a quarter of the round-trip loss);
${\bf M}_{\rm rot}$ is the rotation matrix of the quadratures due to free propagation in vacuum, and it is given by
\begin{equation}
{\bf M}_{\rm rot}=e^{i\,\Omega\,\tau_f}\left[\begin{array}{cc} \cos\Delta_f \tau_f & -\sin\Delta_f \tau_f \\ \sin\Delta_f\tau_f& \cos\Delta_f\tau_f\end{array}\right]
\end{equation}
with $\tau_f=L_f/c$ being the time delay and $L_f$ being the filter cavity length. From these equalities, we
can obtain the corresponding input-output relation between $\bm a$ and $\bm b$. Given small loss: $\eta\ll 1$,
we can keep the input-output relation up to the lowest order of $\epsilon$, and get
\begin{eqnarray}\nonumber
\bm b &=& \left[-\sqrt{R_f}{\bf I}+T_f({\bf I}-\sqrt{R_f} \,{\bf M}_{\rm rot}^2)^{-1} {\bf M}_{\rm rot}^2\right]{\bm a}+
\\\nonumber &&\;\sqrt{{\epsilon}T_f/{4}}({\bf I}-\sqrt{R_f} {\bf M}_{\rm rot}^2)^{-1}({\bf M}_{\rm rot}^2 \bm n_c+{\bf M}_{\rm rot} \bm n_f +{\bf M}_{\rm rot}\bm n_g + \bm n_d)
\\ &&\;-\sqrt{\epsilon R_f/{4}}\,\bm n_a+\sqrt{{\epsilon}/{4}}\,\bm n_b\,.
\end{eqnarray}
The term on the first line gives the cavity response in the lossless case---the quadrature rotation
has a significant frequency dependence; the four terms on the second line are the losses inside the cavity, and they are
amplified by the cavity around the detuning frequency $\Delta_f$; the two terms on the last line are the losses
outside the cavity, and they are not amplified by the cavity response.

To gain an intuitive understanding, we can use the fact that the sideband frequency $\Omega$, the cavity
bandwidth $\gamma_f$ and the detune frequency $\Delta_f$ are much smaller than the free spectral range, namely
\begin{equation}
\Omega \tau_f,\; \gamma_f \tau_f,\; \Delta_f\tau_f \ll 1\, ,
\end{equation}
with $\tau_f \equiv L_f/c$ being the propagation time. We can therefore make a Taylor expansion in terms of
series of these small quantities, and obtain (in the sideband picture):
\begin{eqnarray}\nonumber
b(\Omega)&=&-\frac{\Omega+\Delta_f-i(\gamma_f-\gamma_{\epsilon})}{\Omega+\Delta_f+
i(\gamma_f+\gamma_{\epsilon})}a(\Omega)\\\nonumber
&&-\frac{i\sqrt{\gamma_f\,\gamma_{\epsilon}}}{\Omega+\Delta_f+i(\gamma_f+\gamma_{\epsilon})}
\left[n_c(\Omega)+n_d(\Omega)+n_f(\Omega)+n_g(\Omega)\right]\\
&&-\sqrt{{\epsilon}/{4}}\,n_a(\Omega)+\sqrt{{\epsilon}/{4}}\,n_b(\Omega)\,.
\end{eqnarray}
Here we have defined the effective bandwidth due to loss:
\begin{equation}
\gamma_{\epsilon}\equiv\frac{c\,\epsilon}{L}\,.
\end{equation}
As we can see,  the optical loss has two effects on the performance of the filter cavity: (i) introducing uncorrelated vacuum fluctuation that degrades the sensitivity; (ii) broadening the cavity bandwidth that prevents the use of very short cavity for filtering with desired frequency band. For a round-trip loss of order of tens of ppm, $\epsilon\sim 30\,\rm ppm$, the bandwidth due to loss can be estimated as
\begin{equation}
\gamma_{\epsilon}\approx 2\pi\times 100\,{\rm s}^{-1}\left(\frac{L}{15\,\rm m}\right)\,,
\end{equation}
where the loss limits the bandwidth to be larger than 100Hz for a cavity length around 15m. For the numerical optimization to be discussed, we choose the filter cavity to be of the order of hundreds meter, and in this case, the bandwidth is mainly determined by the transmittance $T_f$ of the input mirror, or equivalently by $\gamma_f$. The loss is mainly important at low frequencies (coherently amplified due to cavity resonance)---it is suppressed by a factor of $\Omega^{-1}$ at high frequencies, and the total effective loss at low frequencies is given by
\begin{equation}
{\cal E}\equiv \frac{4\gamma_f \gamma_{\epsilon}}{(\gamma_f+\gamma_{\epsilon})^2}\approx \frac{\epsilon}{T_{ f}}
\end{equation}
for filter cavity bandwidth $\gamma_f\gg\gamma_{\epsilon}$, which recovers what has been shown in Equation\,\ref{eq:loss_E}.

% some discussion about the effect of parameter variations
\section{Tolerance to parameter variations in the filter cavity}
\label{s:tolerance}
Here we compare input filtering and output filtering in terms of tolerance to parameter uncertainties in the filter cavity.
The outline of this section goes as follows: (i) we first analyze the deviation from the ideal frequency-dependent quadrature rotation due to parameter uncertainties of the filter cavity; (iii) we then show
how these uncertainties influence the sensitivity for both input filtering (frequency-dependent squeezing) and
output filtering (frequency dependent readout).

For a single filter cavity, the output quadratures $(b_1, \, b_2)$ are related to the input quadratures  $(a_1, \, a_2)$
by [cf. Equation\,\ref{eq:filterrotation}]:
\begin{equation}
\left[\begin{array}{c} b_1 \\ b_2\end{array}\right] =
e^{\frac{i(\alpha_+-\alpha_+)}{2}} \left[\begin{array}{cc} \cos \frac{\alpha_++\alpha_-}{2} & - \sin \frac{\alpha_++\alpha_-}{2} \\
 \sin  \frac{\alpha_++\alpha_-}{2} &  \cos  \frac{\alpha_++\alpha_-}{2} \end{array}\right]\left[\begin{array}{c} a_1
 \\ a_2\end{array}\right]\,,
\end{equation}
with $\alpha_{\pm}$ is defined as
\begin{equation}
e^{i\,\alpha_{\pm}}\equiv \frac{i\gamma\mp\Omega-\Delta}{i\gamma\pm\Omega-\Delta}\,,
\end{equation}
where $\Delta$ and $\gamma$ are the detuning frequency and bandwidth of the filter cavity, respectively.
If we have a chain of $N$ filter cavities, the total rotation angle $\phi_{\rm tot}$ is given by
\begin{equation}
\phi_{\rm tot}(\Omega)=\sum_{i=1}^N \frac{\alpha_+^{(i)}(\Omega)+\alpha_-^{(i)}(\Omega)}{2}
\end{equation}
with $\alpha^{(i)}$ from the $i$-th filter cavity. As shown in the Appendix A of \cite{PuCh2002}, by
properly choosing the parameters for each cavity, one can realize any desired frequency-dependent rotation 
angle, as long as $tan(\phi_{\rm tot}(\Omega))$ is a rational function.

Suppose both the detuning frequency and bandwidth have uncertainties $\delta \Delta^{(i)}$ and $\delta \gamma^{(i)}$:
\begin{equation}
\Delta^{(i)} \rightarrow \Delta^{(i)} + \delta \Delta^{(i)}\,, \quad \gamma^{(i)} \rightarrow \gamma^{(i)} +\delta \gamma^{(i)}\,.
\end{equation}
This will induce a change $\delta \phi_{\rm tot}$ in the total rotation angle of
\begin{eqnarray}\nonumber
\delta \phi_{\rm tot} &=&\sum_{i=1}^{N} \frac{\delta \alpha_+^{(i)} +\delta \alpha_-^{i}}{2} \\
                            &=&\sum_{i=1}^N -2\gamma^{(i)}f^{(i)}_+(\Omega)\delta \Delta^{(i)}+2\Delta^{(i)}f^{(i)}_-(\Omega)\delta \gamma^{(i)}\, 
\end{eqnarray}
where
\begin{equation}
f^{(i)}_{\pm}(\Omega)\equiv \frac{\gamma^{(i)2}+\Delta^{(i)2}\pm\Omega^2}{\left[\gamma^{(i)2}+(\Delta^{(i)}-\Omega)^2\right]
\left[\gamma^{(i)2}+(\Delta^{(i)}+\Omega)^2\right]}\,.
\end{equation}
At low frequencies $\Omega\lesssim \gamma^{(i)}$ and for $\gamma^{(i)}\sim \Delta^{(i)}$, we have
\begin{equation}
\delta \phi_{\rm tot}|_{\rm low\;freq.} = \sum_{i=1}^N -\frac{\delta\Delta^{(i)}}{\Delta^{(i)}} + \frac{\delta\gamma^{(i)}}{\gamma^{(i)}}= \sum_{i=1}^N -\frac{\delta\Delta^{(i)}}{\Delta^{(i)}} + \frac{\delta T^{(i)}}{T^{(i)}}\,. 
\end{equation}
Basically, the total rotation angle is just modified by the relative error in the detuning and bandwidth.
In reality, such a relative error can be controlled to a level of $10^{-4}$. This means that
\begin{equation}
\delta\phi_{\rm tot}\sim 10^{-4}N\,.
\label{eq:delta_phi}
\end{equation}

Now we look at how such an error in the quadrature rotation influences the sensitivity for input and output filtering.
We focus on the tuned configuration with the input-output relation shown in Equation \ref{eq:io_tuned}.
For {\it input filtering}, the resulting spectral density is given by (cf. Equations  46-48 in \cite{KLMTV2001}):
\begin{equation}
S_h(\Omega)=\frac{h_{\rm SQL}^2}{2}\left(\frac{1}{\cal K}+{\cal K}\right)(\cosh 2r-\cos[2(\varphi+\Phi)]\sinh 2r).
\end{equation}
with $\Phi={\rm arccot}{\cal K}$. When we choose the optimal squeezing angle
\begin{equation}
\varphi_{\rm opt}=-\Phi(\Omega)=-{\rm arccot}{\cal K}(\Omega)\,,
\end{equation}
we obtain
\begin{equation}
S_h^{\rm opt}(\Omega)=\frac{h_{\rm SQL}^2}{2}\left(\frac{1}{\cal K}+{\cal K}\right) e^{-2r}\,.
\end{equation}
The uncertainties in the parameters of the filter cavity will make $\varphi$ deviate from the optimal squeezing
angle $\varphi_{\rm opt}$, i.e., $\varphi = \varphi_{\rm opt}+\delta \varphi$ resulting in $S_h=S_h^{\rm opt}+\delta S_h$.
Specifically, we have
\begin{equation}\label{eq:dsh_in}
\delta S_h(\Omega)={h_{\rm SQL}^2}\left(\frac{1}{\cal K}+{\cal K}\right) \sinh(2r)\,\delta \varphi^2\,.
\end{equation}
At low frequencies, we have  $\delta S_h(\Omega)|_{\rm low\; freq.}\approx h_{\rm SQL}^2 \sinh(2r)\delta \phi^2/{\cal K}$.

For {\it output filtering}, the spectral density is given by [cf. Equations 56-58 in \cite{KLMTV2001}] with
frequency-independent phase squeezing:
\begin{equation}
S_h(\Omega) = \frac{h_{\rm SQL}^2}{2{\cal K}}\left[e^{-2r}+(\tan\zeta-e^{2r}{\cal K})^2\right]\,.
\end{equation}
When we choose the optimal frequency-dependent homodyne detection angle
\begin{equation}
\zeta(\Omega) ={\rm arctan}({e^{2r}\cal K})\,,
\end{equation}
we obtain the optimal sensitivity that is only limited by the shot noise:
\begin{equation}
S_h^{\rm opt}(\Omega)=\frac{h_{\rm SQL}^2}{2{\cal K}}e^{-2r}\,.
\end{equation}
Similarly, variation of the readout quadrature due to uncertainties of the filter cavity parameters will induce
the following change in the sensitivity: 
\begin{equation}\label{eq:dsh_out}
\delta S_h(\Omega) = \frac{h_{\rm SQL}^2}{{\cal K}}(1+e^{4r}{\cal K}^2)^2 \delta \zeta^2\,.
\end{equation}
Since ${\cal K}\gg 1$ at low frequencies, by comparing Equation\,\ref{eq:dsh_in} and
Equation\,\ref{eq:dsh_out}, we can see that $\delta S_h(\Omega)$ from parameter variations is much larger for the
output filtering case than for input filtering.

\section{Tolerance to anti-squeezing}
\label{sec:antisqz}
Here we analyze the influence of anti-squeezing (an impure squeezed state with impurity from classical
noise during its preparation) on input and output filtering. In contrast to a pure squeezed state, 
the determinant of the noise spectral density matrix for an a state with some anti-squeezing
[cf. Equation\,\ref{eq:squeezematrix}] is:
\begin{equation}
S_{11}S_{22}-S_{12}S_{21}=\xi^2 \ge 1\,,
\end{equation}
This determinant is unity for a pure squeezed state. One example of anti-squeezing is
$S_{11}=\xi^2 e^{2r}$, $S_{12}=S_{21}=0$ and $S_{22}=e^{-2r}$, which means that fluctuations in the phase quadrature
are still the same as the usual pure phase squeezed state but the amplitude quadrature is larger than $e^{2r}$ by a
factor of $\xi^2$. In the ideal case with perfect filter cavities and no loss, such anti-squeezing will not influence
the optimal sensitivity. However, when the loss and the above
mentioned uncertainties in the filter cavity detuning and bandwidth are taken into account, the anti-squeezing 
will play an important role, and in particular, will degrade the low-frequency sensitivity.

Specifically, for the input filtering, we have
\begin{equation}
\delta S_h(\Omega)=\frac{h_{\rm SQL}^2}{2}\left(\frac{1}{\cal K}+{\cal K}\right)(e^{-2r}+e^{2r}\xi^2) \,\delta \varphi^2\,,
\end{equation}
At low frequencies, we can approximate this as $\delta S_h(\Omega)\sim{h_{\rm SQL}^2} {\cal K}e^{2r}\xi^2 \delta \varphi^2/2$.
For the output filtering, we have
\begin{equation}
\delta S_h(\Omega) = \frac{h_{\rm SQL}^2}{{\cal K}}(1+\xi^4 e^{4r}{\cal K}^2)^2 \delta \zeta^2\,.
\end{equation}
where we see that the output filtering is more susceptible to anti-squeezing.

\end{document}